\documentclass[aip,preprint]{revtex4-1}

\usepackage{color}
\usepackage[colorlinks=true,urlcolor=blue,citecolor=red]{hyperref}
\usepackage{graphicx}
\usepackage{amsfonts}
\usepackage{amsmath}

\begin{document}

\title{Comparative analysis of existing models for power-grid synchronization}

\newcommand{\NUPhysicsNICO}{\affiliation{Department of Physics \& Astronomy and Northwestern Institute on Complex Systems, Northwestern University, Evanston, IL 60208, USA}}

\author{Takashi Nishikawa} 
\email{t-nishikawa@northwestern.edu}
\NUPhysicsNICO
\author{Adilson E. Motter}
\NUPhysicsNICO

\begin{abstract}
The dynamics of power-grid networks is becoming an increasingly active area of research within the physics and network science communities. The results from such studies are typically insightful and illustrative, but are often based on simplifying assumptions that can be either difficult to assess or not fully justified for realistic applications. Here we perform a comprehensive comparative analysis of three leading models recently used to study \textit{synchronization dynamics} in power-grid networks---a fundamental problem of practical significance given that frequency synchronization of all power generators in the same interconnection is a necessary condition for a power grid to operate. We show that each of these models can be derived from first principles within a common framework based on the classical model of a generator, thereby clarifying all assumptions involved. This framework allows us to view power grids as complex networks of coupled second-order phase oscillators with both forcing and damping terms. Using simple illustrative examples, test systems, and real power-grid datasets, we study the inherent frequencies of the oscillators as well as their coupling structure, comparing across the different models. We demonstrate, in particular, that if the network structure is not homogeneous, generators with identical parameters need to be modeled as non-identical oscillators in general.  We also discuss an approach to estimate the required (dynamical) parameters that are unavailable in typical power-grid datasets, their use for computing the constants of each of the three models, and an open-source MATLAB toolbox that we provide for these computations (available at \url{https://sourceforge.net/projects/pg-sync-models}).
\end{abstract}

\noindent
{\small\textit{Published in \href{http://dx.doi.org/10.1088/1367-2630/17/1/015012}{New J. Phys. \textbf{17}, 015012 (2015)}}

\maketitle

\section{Introduction}
\label{sec:intro}

In the field of network dynamics, power grids have long served as one of the main model systems motivating the study of synchronization phenomena\cite{Dorogovtsev:2008ly,Arenas:2008yq,Dorfler:2014fk}.
As the field became more mature, an increasing number of researchers have been seeking to apply ideas from past theoretical studies to power grid-specific problems\cite{Mallada:2011lr,PhysRevLett.109.064101,Lozano:2012qy,Dorfler:2013ve,Motter:2013fk,Menck:2014fk,Menck:2014fk,Pecora:2014zr}.
The need for such applications comes from the fact that, despite the extensive engineering literature on power systems, there remains a largely under-explored problem of how the large-scale network structure influences the collective dynamics in power grids.
While previous studies have emphasized the detailed modeling of relatively small test systems, the increasing availability of data processing tools, substantial computing power, and theoretical developments in network synchronization are now making it possible to address large-scale properties of power-grid systems.

A major concern for power grids is the stability of desired states, in particular synchronization stability of the power generators, which is a condition required for their normal operation.
A frequency-synchronous state of $n_g$ power generators is characterized by
\begin{equation}\label{eqn:sync-cond}
\dot\delta_1 = \dot\delta_2 = \cdots = \dot\delta_{n_g},
\end{equation}
where $\delta_i=\delta_i(t)$ is the angle of rotation associated with the $i$-th generator at time $t$.
Studying the stability of synchronous states of an alternating current interconnection against perturbations requires a network model capable of describing the coupled dynamics of power generators.
Different models have been used in different publications in the physics literature, and there has been no comprehensive comparison to clarify how these models are related to each other.
Providing such a comparative analysis is the main focus of this article.

%:fig_modeling
\begin{figure*}
\begin{center}
\includegraphics[height=6in]{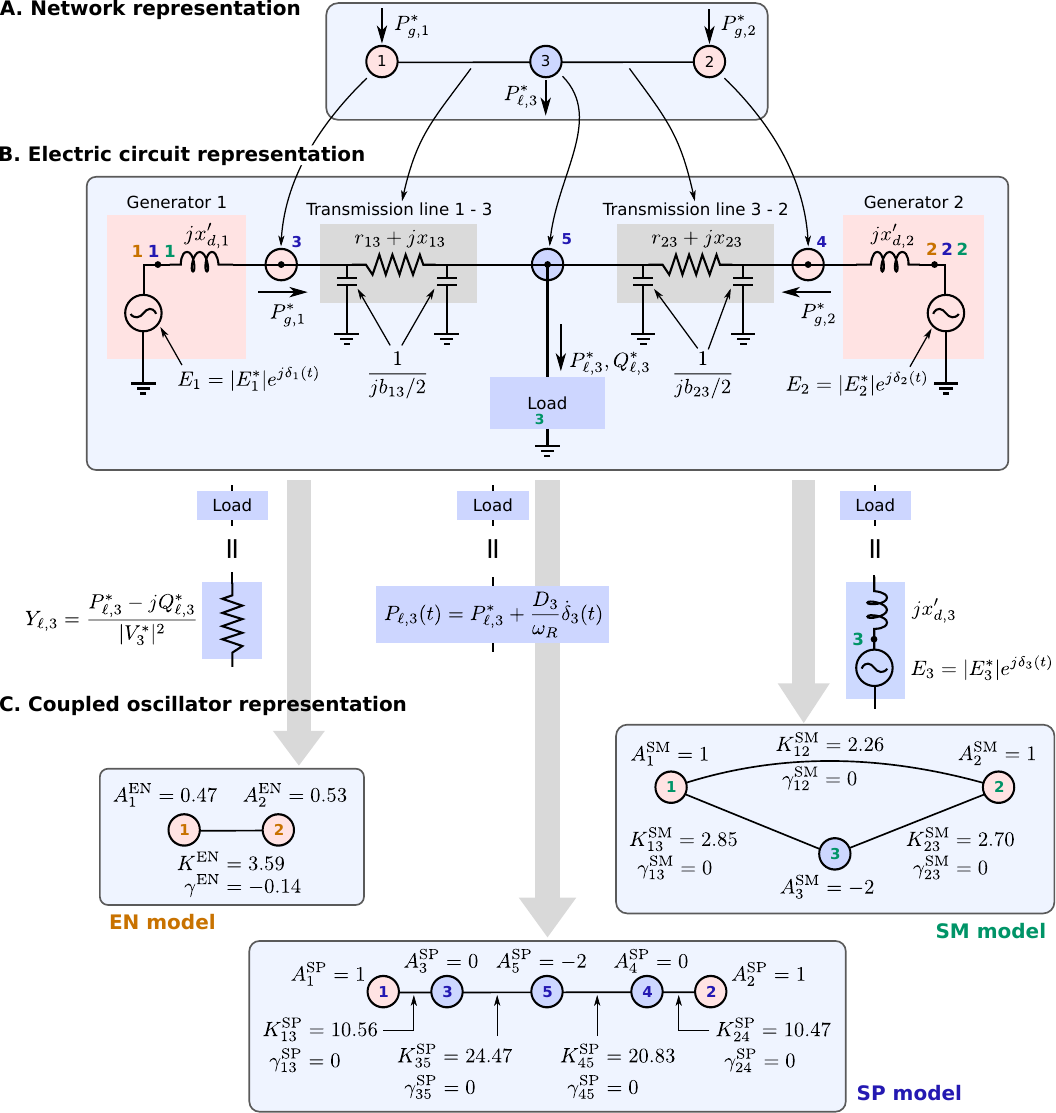}\\
\end{center}
\vspace{-5mm}
\caption{\label{fig:modeling}\baselineskip15pt
Modeling of power-grid network dynamics.
(A) Simple network representation with nodes representing generators or loads, and links representing transmission lines or transformers.
Here we used the example case consisting of two generators (nodes 1 and 2) and one load node (node 3), which is discussed in Section~\ref{sec:two-ten-one-load}.
(B) Representation of the electrical properties of the components in the same network.
There are three possible ways to represent the load node, which are used in three different models.
(C) Representation of the system as a network of coupled oscillators for each choice of load representation in (B).
For the system parameter values given in Section~\ref{sec:two-ten-one-load}, the network dynamics obeys Eq.\,\eqref{model} [or equivalently, Eqs.\,\eqref{eqn:two-gen-EN}, \eqref{eqn:two-gen-SP}, and \eqref{eqn:two-gen-SM} for the EN, SP, and SM model, respectively] with the indicated values of $A_i$, $K_{ij}$, and $\gamma_{ij}$.
Each of the three dynamical models has its own definition of nodes that are different from that used in (A). In (B), these nodes are shown as black dots and indicated by orange, blue, and green indices for the EN, SP, and SM models, respectively.
The same coloring scheme is used for the node indices in (C).
Note that in the SP model the generator terminals are treated as load nodes with zero power consumption (nodes 3 and 4), separately from the generator internal nodes (nodes 1 and 2), leading to the 5-node representation.
}
\end{figure*}

Here, we discuss three leading models, which we refer to as the \emph{effective network} (EN) model\cite{Anderson:1986fk,Motter:2013fk}, the \emph{structure-preserving} (SP) model\cite{Bergen:1981kx}, and the \emph{synchronous motor} (SM) model\cite{springerlink:10.1140/epjb/e2008-00098-8}.
Each model is described as a network of coupled phase oscillators whose dynamics is governed by equations of the form
\begin{equation}\label{model}
\frac{2H_i}{\omega_R}\ddot\delta_i + \frac{D_i}{\omega_R}\dot\delta_i 
= A_i - \sum_{j=1, j\neq i} K_{ij} \sin(\delta_i - \delta_j - \gamma_{ij}),
\end{equation}
where $\omega_R$ is a reference angular frequency for the system, and $H_i$ and $D_i$ are inertia and damping constants characterizing the oscillators, respectively.
The differences in the three models are reflected in the definition of the parameters $A_i$, $K_{ij}$, and $\gamma_{ij}$, as well as in the number of phase oscillators.
The phase angle $\delta_i$ of oscillator $i$ may represent either a generator or load.
While all three models represent the $n_g$ generators as oscillators, they are distinguished mainly by their different modeling approaches for the \emph{loads}, which are representations of individual or aggregated consumers that draw power from specific points in the (transmission) network. 
The EN model represent the loads as constant impedances rather than oscillators, thus focusing on the synchronization of generators as second-order oscillators.
The SP model represents all load nodes as first-order oscillators (i.e., $H_i = 0$),
and each generator is represented by two oscillators, including one for its terminal
(a point connecting the generator to the rest of the network).
The SM model assumes that the loads are synchronous motors that are represented as second-order oscillators. 
Figure\,\ref{fig:modeling} shows a simple example network that illustrates how these differences lead to different network dynamics representations with different number and values of model parameters.
Note that the parameters are denoted in the figure using appropriate superscripts (EN, SP, and SM), which is a convention we will use throughout the article.

The parameter $A_i$, along with $D_i$, determines the $i$-th oscillator's
\textit{inherent frequency}, $\omega_i^* := \omega_R(1 + A_i/D_i)$, which is the equilibrium frequency of oscillator $i$ in the absence of the coupling term [the summation term in Eq.\,\eqref{model}].
For generators, this frequency is typically much higher than the system reference frequency $\omega_R$, as the two terms on the r.h.s.\ of Eq.\,\eqref{model} balance each other in a steady state.
In a realistic setting, however, the instantaneous frequency of a generator is unlikely to actually reach this inherent frequency, since the system operator would take control actions well before the frequency deviates too far from the designated system frequency $\omega_R$.
In addition, the equation of motion for generators and motors we derive below assumes that the frequency remains close to $\omega_R$ and thus would no longer be valid if the frequency deviates too far from $\omega_R$.
Nevertheless, this definition of inherent frequency can be useful in characterizing the nature of individual oscillators and, in fact, is analogous to the one usually used in studying networks of coupled phase oscillators.
Note that ``natural frequency'' is a term commonly used to refer to $\omega^*_i$ in that context, but it has a very different meaning in the power systems literature; see Appendix A.
The parameter $K_{ij} \ge 0$ represents the strength of dynamical coupling between oscillators $i$ and $j$, while $\gamma_{ij}$ represents a phase shift involved in this coupling.
We note that Eq.\,\eqref{model} can be regarded as a second-order analog of the Kuramoto model\cite{kuramoto1984chemical} with arbitrary coupling structure.
Similar forms of second-order equations for coupled oscillators have been used to study synchronization and phase transitions outside the context of power grids\cite{Tanaka1997xd,Acebron1998iogw,Trees2005hok,Ji2013xeo}.

The paper is organized as follows. 
In Sec.\,\ref{sec:steady-state}, we discuss how to characterize and compute the steady state of a power-grid network.
We then turn our attention in Sec.\,\ref{sec:single-gen} to the problem of describing the dynamics of individual generators, deriving the basic equation of motion and describing the electric circuit representation of a generator under the classical model.
In Sec\,\ref{sec:network-dyn}, we derive the EN, SP, and SM models and discuss their differences and similarities.
In Section\,\ref{sec:model-parameters}, we describe how to obtain model parameters required for numerical simulations.
Finally, we discuss in Sec.\,\ref{sec:hetero} an instructive example system and a selection of test systems and real power-grid datasets, focusing particularly on the heterogeneity of the constants $A_i$ and the inherent frequencies as well as on the coupling structure encoded in the constants $K_{ij}$.
We make concluding remarks in Sec.\,\ref{sec:conclusions}.

\section {Steady states of power-grid networks: The power flow equations}
\label{sec:steady-state}

A power grid is a system of electrically coupled devices whose purpose is to deliver power from generators to consumers.
To view the system as a network, we define a \emph{node} to be a point in the system at which power is injected by a generator or extracted by power consumers, or a branching point through which power is redistributed among the transmission lines connected to the point.
A \emph{link} is then defined as an electrical connection between a pair of such nodes and can represent a transmission line or a transformer.
This network representation of the physical structure of a power grid is illustrated in Fig.\,\ref{fig:modeling}A using a simple example of two power-injecting nodes (1 and 2) connected to a single power-consuming node~(3) by two transmission lines.
Transmission lines have impedances, which are represented by complex numbers in general, 
and their parallel conductors have capacitance between them.
In power system analysis, a transmission line is usually represented by the so-called $\pi$ model, in which the two nodes are connected by an impedance, with two capacitors (of equal capacitance) connecting both sides of the impedance to the ground.
This model is illustrated for the transmission line 1--3 in Fig.\,\ref{fig:modeling}B, which has two capacitors with impedance $1/(jb_{13}/2)$, $b_{13}>0$, where we denote the imaginary unit by $j := \sqrt{-1}$.
A transformer is represented by a model in which the (complex-valued) voltages on the two sides of the transformer maintain a constant ratio (generally complex-valued to account for a possible voltage phase shift).
The standard approach\cite{Grainger:1994yo}, which we adopt here, is to represent the parameters of these models for transmission lines and transformers in terms of equivalent admittances (the inverse of impedances).
The structure of the entire physical network can then be represented\cite{Grainger:1994yo} by the (complex-valued) \textit{admittance matrix} $\mathbf{Y}_0 = (\mathbf{Y}_{0ij})$, where an off-diagonal element $Y_{0ij}$ is the negative of the admittance between nodes $i$ and $j\neq i$ and a diagonal element $Y_{0ii}$ is the sum of all admittances connected to node $i$ (including the shunt admittances to the ground, which are parts of the models for transmission lines and transformers).

The operating condition of a power grid can be characterized by the distribution of power flow through the physical network of transmission lines and transformers.
In an alternating current grid, power 
is represented as
a complex number, 
whose real and imaginary components, $P_i$ and $Q_i$, are called \textit{active}
and \textit{reactive} power, respectively.
Alternatively, the power flow state can be characterized uniquely by the complex voltage $V_i$ (with magnitude $|V_i|$ and phase angle $\phi_i$, so that $V_i = |V_i|e^{j\phi_i}$) and the complex power $P_i + jQ_i$ injected into each node $i$ in the network.
Given the locations and parameters of power generators and loads, as well as the parameters of the network summarized in the admittance matrix $\mathbf{Y}_0$, fundamental laws of physics---the Kirchhoff's laws---determine the power flow state of the system.
The laws can be used to derive the so-called \emph{power flow equations}\cite{Grainger:1994yo}, which provide the foundation for steady-state power system analysis:
\begin{align}
P_i &= \sum_{j=1}^n |V_i V_j Y_{0ij}|\sin(\phi_i - \phi_j - \gamma_{0ij}), \quad i = 1,\ldots,n, \label{pf-eqn-p}\\
Q_i &= - \sum_{j=1}^n |V_i V_j Y_{0ij}|\cos(\phi_i - \phi_j - \gamma_{0ij}), \quad i = 1,\ldots,n,\label{pf-eqn-q}
\end{align}
where the complex admittances are represented in polar form as $Y_{0ij} = |Y_{0ij}|e^{j\alpha_{0ij}}$, we define $\gamma_{0ij} := \alpha_{0ij} - \pi/2$, and $n$ is the number of nodes.
To solve this set of $2n$ nonlinear equations for all $4n$ quantities that determine the power flow state of the system, we require that two real quantities per node are given as parameters.  The usual assumptions are:
\begin{itemize}
\item If node $i$ is a generator node, the values of $P_i$ and $|V_i|$ are given as $P_i = P_{g,i}^*$ and $|V_i| = |V_i^*|$.
This is reasonable because in real power grids the generators are scheduled to produce constant active power at a given time, and constant voltage magnitude is usually maintained by voltage regulators.
Usually, one generator node is chosen to be a reference node, for which $\phi_i$ is set to zero instead of specifying the value of $P_i$.
For the steady-state analysis, we need at least one node with unspecified $P_i$ because the total power generation cannot be known \textit{a priori} even when the total power consumption is known, since the difference, which equals the power lost in the transmission lines and transformers, depends on the power flow state.
\item If node $i$ is a load node, the values of $P_i$ and $Q_i$ are given as
$P_i = -P_{\ell,i}^*$ and $Q_i = -Q_{\ell,i}^*$.
For positive active and reactive power consumptions, $P_i$ and $Q_i$ take negative values.
These values can be determined from historical data, load forecast, or measurement~\cite{Anderson:1986fk,Grainger:1994yo}.
\end{itemize}
Given the parameter values and the admittance matrix $\mathbf{Y}_0$, the nonlinear equations~\eqref{pf-eqn-p} and \eqref{pf-eqn-q} are solved numerically, which can be performed by a variety of software, including MATLAB-based, freely available packages such as Power System Toolbox (PST)~\cite{chow1992toolbox} and MATPOWER~\cite{matpower}.
In the following, we will denote the values of the state variables for a power flow solution as $P^*_i, Q^*_i, V^*_i$, and $\phi^*_i$.

\section{Dynamics of individual generators within the network}
\label{sec:single-gen}

\subsection{Mechanical representation --- The swing equation}
\label{sec:swing}

Many components of a power grid, such as generators and loads, are dynamic and the state of the component can change in time.
A dynamical model of the grid is obtained when the two parameters given for each node to determine the power flow state are replaced by a set of differential equations that, together with the power flow equations [Eqs.~\eqref{pf-eqn-p} and \eqref{pf-eqn-q}], allow the determination of the state of the system as a function of time.
Instead of the constant power injection and voltage magnitude at a generator node, one can write an equation of motion describing the dynamics of the generator rotor.

Most generators in today's power grids have a rotating mass (a rotor) that is driven by mechanical torque to generate electrical power.
There exist models with various levels of sophistication and complexity, but in all such models, the rotational dynamics of the rotor is ultimately governed by a fundamental law of physics: the Newton's second law.
The electrical output from such a generator is coupled to other generators in the grid through a network of transmission lines, transformers, and other devices, which serve to transport the generated power to consumers.
The power demanded by the network is felt by the generator's rotor as an electrical torque, which is usually a decelerating torque.
It is the balance between the mechanical input power and the electrical output power that determines the dynamics of the rotor.

To derive the equation of motion governing the dynamics of such a generator, we set the rate of change of the angular momentum of the rotor equal to the net torque acting on the rotor:
\begin{equation}
J \ddot\delta = \tilde{T}_m - D_m\omega - \frac{1}{R} \Delta\omega - D_e \Delta\omega - T_e, 
\end{equation}
where $J$ is the moment of inertia in kg$\cdot$m$^2$, $\delta$ is the angle of the rotor relative to a frame rotating at the reference frequency $\omega_R$ in rad/s, $\tilde{T}_m$ is the mechanical torque in N$\cdot$m driving the rotor, $D_m$ is the damping coefficient in N$\cdot$m$\cdot$s for mechanical friction, $\omega$ is the angular frequency of the rotor in rad/s, $R$ is the regulation parameter in rad/N$\cdot$m$\cdot$s characterizing the proportional frequency control by a governor, $\Delta\omega := \omega - \omega_R$ is the frequency deviation, $D_e$ is the damping coefficient in N$\cdot$m$\cdot$s for the electrical effect of the generator's damper windings, and $T_e$ is the typically decelerating torque in N$\cdot$m due to electrical load in the network.
Noting that $\dot\delta = \omega - \omega_R = \Delta\omega$, we can rewrite the equation as
\begin{equation}
J \ddot\delta + \tilde{D}\dot\delta= T_m - T_e, 
\end{equation}
where $\tilde{D} := D_m + D_e + 1/R$ and $T_m := \tilde{T}_m - D_m\omega_R$ is the net mechanical torque, accounting for frictional loss at the reference frequency.
Multiplying both sides by $\omega$ and using the fact that torque in N$\cdot$m multiplied by angular velocity in radians per second gives power in watts, the equation can be written in terms of power: 
\begin{equation}
J\omega_R \ddot\delta + \tilde{D}\omega_R \dot\delta = \frac{\omega_R}{\omega}(T_m \omega - T_e \omega) \approx \tilde{P}_m - \tilde{P}_e,
\end{equation}
where we define $\tilde{P}_m := T_m \omega$ and $\tilde{P}_e := T_e \omega$, and we assumed  the factor $\omega_R/\omega$ to be nearly equal to one, i.e., that the generator's frequency $\omega$ is close to the reference frequency $\omega_R$.

This approximation, which can be formalized using singular perturbation analysis~\cite{sauer1998power},
is valid and considered appropriate for studying power system stability, where the key question is whether the system desynchronizes after a perturbation from a steady-state operation near the reference frequency.
If, for example, a disturbance leads to even 1\% deviation of a generator's frequency from the 60 Hz reference frequency for a period of just one second, it would lead to an increase in angle difference equivalent to more than half a rotation.
This much of angle deviation is highly likely to cause loss of synchronization in practical situations~\cite{Anderson:1986fk}.
While the approximation $\omega \approx \omega_R$ is widely accepted in this context, a more detailed electromechanical model that does not require this approximation is also available~\cite{Caliskan2014slgd}.

We now divide both sides of the equation by the rated power $P_R$ (used as a reference)
to make $\tilde{P}_m$ and $\tilde{P}_e$ \textit{per unit} (p.u.) quantities, which is a normalization procedure commonly used in power systems studies.
The factor $J\omega_R$ then becomes $2H/\omega_R$, where we defined the inertia constant $H := W/P_R$ (in seconds) and the kinetic energy of the rotor $W := \frac{1}{2}J\omega_R^2$ (in joules).
The factor $\tilde{D}\omega_R$ becomes $D/\omega_R$, where we defined the combined damping coefficient as $D := \tilde{D}\omega_R^2/P_R$ (in radians). 
We then obtain what is known in the power systems literature as the \emph{swing equation}~\cite{Anderson:1986fk,Grainger:1994yo,sauer1998power}:
\begin{equation}\label{swing-eqn}
\frac{2H}{\omega_R}\ddot\delta + \frac{D}{\omega_R}\dot\delta = P_m - P_e,
\end{equation}
which we use here as the fundamental equation of motion for a generator.
The term $P_m$ represents the net mechanical power input to the rotor, while $P_e$ represents the electrical power demanded by the rest of the network and includes terms that depend explicitly on $\delta$ and the state variables of the other generators and loads in the network.

\subsection{Electric circuit representation --- The classical model}
\label{sec:classical-model}

In general, both $P_m$ and $P_e$ have nonlinear dynamics, which may need to be accounted for in high-accuracy simulations that power system engineers require.
There has been substantial effort in the power systems literature to model the dynamical behavior of the generator's internal magnetic flux, which affects the electrical power output $P_e$~\cite{Anderson:1986fk,pai1989energy,Grainger:1994yo,Chu:1999kx,Chu:2005uq,Caliskan2014slgd}.
One may also need to include the nonlinear dynamics of the governor that controls the generator frequency and the excitation system that controls the voltage magnitude at the generator terminal.
Here we aim to study power grids as a complex dynamical system and focus on the way in which the network structure influences the synchronization dynamics of generators in simplest settings that are nevertheless realistic.
For this purpose, we now describe a model of a generator used commonly in the engineering literature, particularly in theoretical studies, and forms a basis of all three network models we present below.
In this so-called \emph{classical model}, a generator is represented as a voltage source with constant voltage magnitude $|E|$ connected to the terminal node through a reactance $ x'_d > 0$ (see Fig.\,\ref{fig:modeling}B).
The phase angle of the voltage source is assumed to be equal to the rotor's rotational angle and thus denoted by $\delta$.
In addition, the mechanical power input $P_m$ to the rotor in Eq.~\eqref{swing-eqn} is assumed to be constant.
Constant voltage magnitude and mechanical power are approximations that are valid for short-term dynamics.
For analysis of transient stability after a disturbance, the approximation is considered valid for simulation of the ``first swing'' of the resulting oscillation of $\omega$ or $\delta$, which typically covers a time period on the order of one second\cite{Anderson:1986fk}.
The classical model sometimes refers to the one in which damping is ignored [i.e., $D=0$ in Eq.~\eqref{swing-eqn}], but here we consider damping effects explicitly, as they can have non-negligible effect on the stability of steady-state power grid operation and can potentially be used as tunable parameters for optimizing the stability\cite{Motter:2013fk}.

In a real power grid under a steady-state condition, the system frequency is continuously monitored and is in fact actively controlled (by adjusting $P_m$ of its generators) to stay close to the reference frequency of the system.
Because of this, in power system stability analysis, it is usually assumed that all generators are initially synchronized at the reference frequency $\omega_R$ in a steady state.
Then, for a given steady state in which the generator is delivering active power $P_g^*$ through its terminal, we have $\ddot\delta = \dot\delta = 0$ and $P_m = P_e = P_g^*$.
This value of $P_m$ is then held constant when studying stability using the swing equation and the classical model of the generator.

Regardless of whether the grid is under steady-state condition or not, the expression for the electrical output power is given by the so-called power-angle equation, $P_e = \frac{|E^* V|}{x'_d} \sin(\delta - \phi)$,
where the complex voltage $V = |V| e^{j\phi}$ at the terminal has generally time-dependent magnitude and phase, and $E = |E^*| e^{j\delta}$ is the voltage at the internal node defined to be a point between the constant-magnitude voltage source (with time-dependent phase) and the transient reactance (see Fig.\,\ref{fig:modeling}B).
Equation~\eqref{swing-eqn} then becomes
\begin{equation}\label{swing-eqn-cls}
\frac{2H}{\omega_R}\ddot\delta + \frac{D}{\omega_R}\dot\delta = P^*_g - \frac{|E^* V|}{x'_d} \sin(\delta - \phi),
\end{equation}
While $H$, $D$, and $x'_d$ are constants that characterize physical and electrical properties of the specific generator, $P^*_g$ and $|E^*|$ are parameters that depend on the steady-state distribution of power flow across the network.
By eliminating the current $I$ from the expression for the complex power, $P^*_g + j Q^*_g = V\bar{I}$ (where the bar denotes the complex conjugate), and the Ohm's law, $j x'_d I = E - V$, applied to the transient reactance, 
the steady-state internal voltage magnitude $|E^*|$ can be calculated as
\begin{equation}\label{eqn-E}
|E^*|^2 = \left( \frac{P^*_g x'_d}{|V^*|} \right)^2 
+ \left(|V^*| + \frac{Q^*_g x'_d}{|V^*|}\right)^2,\\
\end{equation}
where $Q_g^*$ is the reactive power injected by the generator into the terminal and $|V^*|$ is the terminal voltage magnitude in the steady state.
We thus see that the state of the terminal node given by $P_g^*$, $Q_g^*$, and $|V^*|$ determines some of the parameters required to simulate generator dynamics using Eq.~\eqref{swing-eqn-cls}.
This is a crucial point that, as we will show below, has significant implications for the modeling of a power grid as a network of coupled oscillators.
In a typical stability analysis using the classical model, $P_g^*$ and $|V^*|$ are given as input data, and $Q_g^*$ is obtained by solving the power flow equations~\eqref{pf-eqn-p} and \eqref{pf-eqn-q} for the entire network, given necessary input data at other nodes.
The power flow solution also provides values of the phase angles, $\delta^*$ and $\phi^*$, in the steady state, which can be used as the initial condition for Eq.~\eqref{swing-eqn-cls}.
Note, however, that the equation is not closed, as the dynamics of $\phi = \phi(t)$ and $V = V(t)$ are not yet specified.
It is mainly the phase angle $\phi$ that serves as the medium through which this generator is coupled to the rest of the system.

\section{Collective dynamics of generators in power-grid networks}
\label{sec:network-dyn}

The dynamical state of the network is characterized by four functions of time for each node: $P_i$, $Q_i$, $|V_i|$, and $\phi_i$.
The power flow equations~\eqref{pf-eqn-p} and \eqref{pf-eqn-q}, being valid at each instant of time for time-varying variables, provide two out of the four required equations per node for uniquely determining the dynamical state of the system.
We thus need two additional equations/assumptions at each node.
For a generator node, the swing equation effectively provides one equation, and the usual practice is to make the additional assumption that the reactive power $Q_g$ injected by the generator at its terminal node is constant over short time scales and equals its steady-state value $Q_g^*$, which is computed from the (static) power flow equations.
We would then have a complete set of equations that allows us to determine all four state variables of the generator, $|E|$, $\delta$, $P_g$, and $Q_g$, as a function of time, given the initial condition and the state variables (for all $t$) at all the other nodes.
Note that $E = |E|e^{j\delta}$ is related to the terminal voltage $V = |V|e^{j\phi}$ through the transient reactance at all times.
Of the four generator variables, $\delta$ directly relates to Eq.\,\eqref{eqn:sync-cond} and thus is the most relevant for the problem of synchronization stability.

Supplying the full set of equations for load nodes requires modeling of static and/or dynamic behavior of the loads.
Load modeling is a hard problem because the power consumption at a node in a transmission network is typically an aggregate of a large number of loads of a wide variety of types and sizes, each of which can be time dependent and influenced by human activity.
As a consequence, a number of different models have been proposed and used in the power system literature.
As mentioned above, it is desirable to have a simple but realistic model that can clarify the role of network structure in the problem of power-grid synchronization stability.
We thus provide in this article a comparative analysis of three approaches that can be used to recast the problem as that of synchronization in complex networks of coupled oscillators.
These approaches lead to the three different models, the EN, SP, and SM models, which are all described by Eq.\,\eqref{model} but with different values and interpretations for the parameters $A_i$, $K_{ij}$, and $\gamma_{ij}$.

\subsection{The effective network model}
\label{sec:eff-net}

The dynamical interaction between a pair of generators is mediated by the paths of transmission lines connecting the generators and is expressed as the sinusoidal dependence on the terminal voltage phase in Eq.~\eqref{swing-eqn-cls}.
While the nature of the coupling between generators is ultimately shaped by the structure of the transmission network and the distribution of loads, it would be insightful and useful if one could represent the coupling in a single term that depends on the state variables of the generators.
This can indeed be achieved by modeling loads as constant impedances and reducing the physical network to what we call the \emph{effective network} of interactions between generators. 
Since the model is derived through this reduction process, it is also called the network-reduction model or network-reduced model in the literature.

In order to see how the reduction process leads to the effective network, which then determines the coupling constants $K_{ij}$ governing the network dynamics in Eq.\,\eqref{model},
let $n_g$ and $n_\ell$ denote the number of generator terminal nodes and load nodes, respectively, so that $n = n_g + n_\ell$.
In addition to these $n$ nodes, we define an additional node to be a point between the internal transient reactance and the constant voltage source for each generator (see the two generators depicted in Fig.\,\ref{fig:modeling}B), making the total number of nodes $N := 2n_g + n_\ell = n_g + n$.
By suitable re-indexing, we may assume that $i=1,\ldots,n_g$ correspond to the generator internal nodes, $i=n_g+1, \ldots, 2n_g$ to generator terminal nodes, and $i=2n_g+1,\ldots,N$ to load nodes.
Define $\mathbf{Y}_d$ as the $n_g \times n_g$ diagonal matrix whose diagonal elements are the admittances $(j x'_{d,1})^{-1},\ldots, (j x'_{d,n_g})^{-1}$ of the generator transient reactances.
We write the admittance matrix $\mathbf{Y}_0$ for the physical network as
\begin{equation}
\mathbf{Y}_0 =
\begin{pmatrix}
\mathbf{Y}_0^{gg} & \mathbf{Y}_0^{g\ell} \\
\mathbf{Y}_0^{\ell g} & \mathbf{Y}_0^{\ell\ell}
\end{pmatrix}
\end{equation}
where $\mathbf{Y}_0^{gg}$, $\mathbf{Y}_0^{g\ell}$, $\mathbf{Y}_0^{\ell g}$, and $\mathbf{Y}_0^{\ell\ell}$ are the four blocks resulting from separating the first $n_g$ rows (columns) from the last $n_\ell$ rows (columns).

The active power $P^*_{\ell,i}$ and reactive power $Q^*_{\ell,i}$ consumed at load node $i$ (and possibly also at a generator terminal node) in a steady state is represented by an equivalent impedance to the ground\cite{Anderson:1986fk} whose admittance is $Y_{\ell, i} = (P^*_{\ell,i} - jQ^*_{\ell,i})/|V^*_i|^2$, where $|V^*_i|$ is the voltage magnitude at the node.
The constants $P_{\ell,i}^*$ and $Q_{\ell,i}^*$ are part of the input data for Eqs.~\eqref{pf-eqn-p} and \eqref{pf-eqn-q} to solve for a power flow solution, which provides $|V_i^*|$.
The constant admittance $Y_{\ell, i}$ is computed from these steady-state values and added to the corresponding diagonal components of $\mathbf{Y}_0^{gg}$ and $\mathbf{Y}_0^{\ell\ell}$ to obtain $\tilde{\mathbf{Y}}_0^{gg}$ and $\tilde{\mathbf{Y}}_0^{\ell\ell}$.
This representation requires the assumption that the power demand is constant.  We thus consider the dynamics of the system on time scales that are short enough for the validity of the assumption, but long enough to address the problem of synchronization. 
We can now define an $N \times N$ admittance matrix $\mathbf{Y}'_0 = (Y'_{0ij})$ that includes the links representing the transient reactances connecting the generators' internal and terminal nodes, as well as the equivalent impedances for the loads:
\begin{equation}\label{ext-adm-mat}
\mathbf{Y}'_0 := \left(
\begin{array}{c|c|c}
\mathbf{Y}_d & -\mathbf{Y}_d\hspace{3mm} & \mathbf{0}\hspace{1mm} \\
\hline
-\mathbf{Y}_d\hspace{3mm} & \tilde{\mathbf{Y}}_0^{gg} + \mathbf{Y}_d & \hspace{2mm}\mathbf{Y}_0^{g\ell} \\
\hline
\mathbf{0}\hspace{1mm} & \mathbf{Y}_0^{\ell g} & \hspace{2mm}\tilde{\mathbf{Y}}_0^{\ell\ell}
\end{array}
\right),
\end{equation}
where $\mathbf{0}$ denotes matrices (of appropriate sizes) whose elements are all zeros.

Now let $\mathbf{V}^g$, $\mathbf{V}^t$, and $\mathbf{V}^\ell$ be the vectors of node voltages for generator internal nodes, generator terminal nodes, and load nodes, respectively, and stack them vertically in that order to form the voltage vector $\mathbf{V}$.
Also let $\mathbf{I}^g$ be the vector of currents injected into the system at the generator internal nodes.
Because the loads are modeled as constant impedances to the ground, all nodes have zero injection currents except for the generator internal nodes.
The Kirchhoff's current law can then be written in the form $\mathbf{I} = \mathbf{Y}'_0 \mathbf{V}$, or equivalently
\begin{equation}\label{eqn:kirchhoff}
\begin{pmatrix}
\mathbf{I}^g \\
\mathbf{0}\\
\mathbf{0}
\end{pmatrix} = \left(
\begin{array}{c|c|c}
\mathbf{Y}_d & -\mathbf{Y}_d\hspace{3mm} & \mathbf{0}\hspace{1mm} \\
\hline
-\mathbf{Y}_d\hspace{3mm} & \tilde{\mathbf{Y}}_0^{gg} + \mathbf{Y}_d & \hspace{2mm}\mathbf{Y}_0^{g\ell} \\
\hline
\mathbf{0}\hspace{1mm} & \mathbf{Y}_0^{\ell g} & \hspace{2mm}\tilde{\mathbf{Y}}_0^{\ell\ell}
\end{array}
\right)
\begin{pmatrix}
\mathbf{V}^g \\
\mathbf{V}^t \\
\mathbf{V}^\ell
\end{pmatrix}.
\end{equation}
The system is then converted to
$\textbf{I}^g = \textbf{Y}\,\textbf{V}^g$ by eliminating
$\textbf{V}^t$ and $\textbf{V}^\ell$, a procedure known as Kron reduction\cite{Grainger:1994yo}, where the resulting {\it effective} admittance matrix $\mathbf{Y}^\text{EN} = (Y^\text{EN}_{ij})$ is defined by
\begin{align}\label{eq:eff-adm}
\mathbf{Y}^\text{EN} &:= \mathbf{Y}'(\mathbf{1} + \mathbf{Y}_d^{-1} \mathbf{Y}')^{-1}, \quad \text{where}\quad 
\mathbf{Y}' := \tilde{\mathbf{Y}}_0^{gg} - \mathbf{Y}_0^{g\ell}
(\tilde{\mathbf{Y}}_0^{\ell\ell})^{-1} \mathbf{Y}_0^{\ell g},
\end{align}
where $\mathbf{1}$ denotes the $n_g \times n_g$ identity matrix.
The inverse of $\mathbf{Y}_d$ always exists, while the matrix $(\mathbf{1} + \mathbf{Y}_d^{-1} \mathbf{Y}')^{-1}$ exists when $x'_{d,i}$ are small enough. 
The existence of the matrix $(\mathbf{Y}_0^{\ell\ell})^{-1}$
follows from the assumed uniqueness of the voltage vectors (with 
respect to a reference voltage). 
The symmetric $n_g \times n_g$ matrix $\mathbf{Y}^\text{EN}$ defines an electrically equivalent network in which generator internal nodes $i$ and $j\neq i$ are connected by an effective admittance $-Y^\text{EN}_{ij}$.
Note that this notion of effective admittance is distinct from the inverse of the (two-point) effective impedance used in AC circuit theory~\cite{Tzeng2006lsg}, but it is more suitable for our purpose here since it captures the idea that dynamical interactions between generators are determined by an effective network of admittances connecting the generators.
Also note that a similar effective network can be derived in the case of constant current injections at load nodes and generator terminal nodes [nonzero current vectors $\mathbf{I}^t$ and $\mathbf{I}^\ell$ replacing the zero entries on the left hand side of Eq.\,\eqref{eqn:kirchhoff}].

Accounting for the transient reactances $x'_{d,i}$ is important, since they are typically not small~\cite{Grainger:1994yo} for real generators producing small power and vary widely from generator to generator.
This is because the values need to be expressed in p.u.\ with respect to the system base (a common set of reference units for the system) when writing the admittance matrices $\mathbf{Y}_d$ and $\mathbf{Y}'_0$.
If instead the values were expressed in p.u.\ with respect to the rated power $P_R$ of the generator, they tend to lie in a relatively narrow range\cite{Grainger:1994yo}.
A consequence of having $x'_{d,i}>0$ for all $i$ is that the effective network represented by $\mathbf{Y}^\text{EN}$, which is obtained after eliminating all nodes except for the generator internal nodes, has the topology of a complete graph ($Y_{ij}\neq 0$ for all $i$ and $j$).
This follows from a general property of Kron reduction that two nodes are connected in the reduced network if and only if the two nodes are connected in the original network by a path in which all intermediate nodes are eliminated by the reduction process\cite{Dorfler:2011fa}.
If one neglects the transient reactances and sets $x'_{d,i} = 0$, we see from Eq.\,\eqref{eq:eff-adm} that we would have $\mathbf{Y}^\text{EN} = \mathbf{Y}'$.
The matrix $\mathbf{Y}'$ can be shown to equal the admittance matrix between the generator terminal nodes obtained by eliminating the load nodes through Kron reduction, which may or may not have the topology of a complete graph depending on the location of the load nodes.

Using the effective admittance matrix $\mathbf{Y}^\text{EN}$, the single sinusoidal coupling term in Eq.~\eqref{swing-eqn-cls} can be replaced by an expression for $P_e$ that comes from a power balance equation equivalent to Eq.~\eqref{pf-eqn-p} with $|V_i|$ replaced by $|E^*_i|$, $\mathbf{Y}_0$ by $\mathbf{Y}^\text{EN}$, and $\phi_i$ by $\delta_i$:
\begin{equation}
P_{e,i} = \sum_{j=1}^{n_g} |E^*_i E^*_j Y^\text{EN}_{ij}|\cos(\delta_j - \delta_i + \alpha^\text{EN}_{ij}),
\end{equation}
where $Y^\text{EN}_{ij} = |Y^\text{EN}_{ij}| \exp(j\alpha^\text{EN}_{ij})$ and all quantities must be expressed in p.u.\ with respect to the system base.
This can be used to show that writing the swing equation~\eqref{swing-eqn} for each generator leads to an equation of the same form as Eq.~\eqref{model}, and the EN model is thus given by
\begin{equation}
\begin{split}
&\frac{2H_i}{\omega_R}\ddot\delta_i + \frac{D_i}{\omega_R}\dot\delta_i 
= A^\text{EN}_i - \sum_{j=1,j\neq i}^{n_g} K^\text{EN}_{ij} \sin(\delta_i - \delta_j - \gamma^\text{EN}_{ij}),
\quad i = 1,\ldots,n_g,\\
&A^\text{EN}_i := P_{g,i}^* - |E^*_i|^2 G^\text{EN}_{ii},\quad
K^\text{EN}_{ij} := |E^*_i E^*_j Y^\text{EN}_{ij}|,\quad
\gamma^\text{EN}_{ij} := \alpha^\text{EN}_{ij} - \frac{\pi}{2},
\end{split}
\end{equation}
where $G^\text{EN}_{ii}$ is the real part of the complex admittance $Y^\text{EN}_{ii}$. 
Notice that the number of phase oscillators in Eq.\,\eqref{model} in this case is $n_g$, since generators are the only dynamical elements over the relevant time scales and each generator is described by a single variable $\delta_i$.
The internal voltage magnitude $|E^*_i|$ is a constant computed from a given steady-state power flow by applying Eq.~\eqref{eqn-E} to each generator.
Based on the same reasoning we used at the single-generator level in Sec.~\ref{sec:classical-model}, the constant net mechanical power input to the generators is equal to the steady-state electrical power output $P_{g,i}^*$.
It can be seen from this equation that the parameter $A^\text{EN}_i$ [which determines the inherent frequency $\omega^*_i = \omega_R(1 + A^\text{EN}_i/D_i)$] and the coupling strength $K^\text{EN}_{ij}$ depend not only on the structure of the transmission network expressed in $\mathbf{Y}_0$, but also on the steady-state power flow over the network through the values of $P_{g,i}^*$, $P_{\ell,i}^*$, $Q_{\ell,i}^*$, and $V_i^*$, which affect $\mathbf{Y}^\text{EN}$ and $|E^*_i|$.
The effective admittances in $\mathbf{Y}^\text{EN}$ for realistic systems have imaginary parts that are mostly positive and much larger than the real parts\cite{Motter:2013fk}, indicating that inductive reactances are the dominant components of the effective network.
This leads to phase shifts $\gamma^\text{EN}_{ij} \approx 0$, and we see that the individual coupling terms in Eq.~\eqref{model} tend to keep the angle differences between generators small, reflecting an inherent tendency of connected generators to synchronize with each other.
This justifies the use of the assumption that all transmission lines are lossless (i.e., their impedances have no real components) and inductive (i.e., their impedances have positive imaginary components), since it implies that Re$(Y^\text{EN}_{ij}) = 0$, Im$(Y^\text{EN}_{ij}) > 0$ and $\gamma^\text{EN}_{ij} = 0$ for all $i \neq j$.
Note, however, that this assumption does not imply that the diagonal components $Y^\text{EN}_{ii}$ have zero real components ($G^\text{EN}_{ii} = 0$), and thus does not imply that $A^\text{EN}_i = P_{g,i}^*$.

In a recent publication\cite{Motter:2013fk}, we derived a master stability function\cite{Pecora:1998zp} for the EN model,  which reduces the problem of synchronization stability against small perturbations to a simple spectral condition that involves both the effective network and the synchronous state.
Based on this condition, we identified a systematic adjustment of generator parameters that enhances the synchronization stability and demonstrated the effectiveness of this approach using a selection of test systems and real power-grid datasets.
Sufficient conditions for synchronization stability against larger perturbations have also been derived for the EN model, one in terms of the maximum voltage phase difference in an approximation of the corresponding power flow solution\cite{Dorfler:2013ve} (assuming lossless transmission lines), and another in terms of the minimum combined coupling strength (accounting for $K_{ij}$, $\gamma_{ij}$, and $D_i$) among all node pairs\cite{fd-fb:09z}.
A similar sufficient condition has also been derived for an extension of the EN model that incorporates a second-order model for the frequency control of individual generators\cite{Wu:2014lr}.
It has been found\cite{Menck:2014fk} that certain structural motifs can cause the loss of synchronization stability in a special case of the EN model which assumes that all nodes have generators with identical parameters and negligible $x'_{d,i}$, all transmission lines are lossless and have equal reactance, and half of the generators (randomly chosen) have $A^\text{EN}_i=P$ and the other half have $A^\text{EN}_i=-P$ for some constant $P>0$.
Under these assumptions, it can be shown that $\mathbf{Y}^\text{EN} = \mathbf{Y}'$ has the same topology as the network of transmission lines.
This simplified version of the EN model corresponds to a power grid that is homogeneous in all aspects, except that the locations of power generators and consumers can be heterogeneous and the network topology can be arbitrary.

\subsection{The structure-preserving model}
\label{sec:struct-preserving}

A different approach is to seek to describe the dynamic behavior of the loads, which can lead to a model of the system that retains the structure of the physical network of transmission lines connecting the load nodes.
In a steady state, the active and reactive power injected into load node $i$ are constant: $P_{\ell,i} = P_{\ell,i}^*$ and $Q_{\ell,i} = Q_{\ell,i}^*$.
The equivalent admittance we used for the loads in the EN model above means that the dynamic behavior of the power consumption by the load is described by a function of the time-dependent node voltage magnitude: $P_{\ell,i} = G_{\ell,i}|V_i(t)|^2$ and $Q_{\ell,i} = B_{\ell,i}|V_i(t)|^2$, where $G_{\ell,i}$ and $B_{\ell,i}$ are the real and imaginary parts of the constant admittance $Y_{\ell,i}$.
In general, the power consumption can depend nonlinearly on both the node voltage magnitude $|V_i(t)|$ and phase frequency $\dot\phi_i(t)$.
The SP model uses an alternative model for the dynamic behavior of active power consumption, $P_{\ell,i} = P_{\ell,i}^* + \frac{D_i}{\omega_R} \dot\delta_i(t)$, where $D_i > 0$ is a constant and $\delta_i(t) := \phi_i(t) - \omega_R t$ is the voltage phase relative to a common reference frame rotating at the synchronous frequency\cite{Bergen:1981kx}.
This model results from the linearization of the power-frequency relation, assuming constant node voltage magnitude $|V_i| = |V_i^*|$ and small deviation $\dot\delta_i$ from the steady-state frequency.
The reactive power consumption is assumed to be constant, $Q_{\ell,i} = Q_{\ell,i}^*$.
This load model leads to the equation of the same form as Eq.~\eqref{model}, where we set $H_i = 0$ and re-interpret $D_i$ as the linear coefficient in the frequency dependence of the active power consumption.
The structure of the network connecting the generators and loads is given by an $N \times N$ matrix similar to $\mathbf{Y}'_0$ in Eq.~\eqref{ext-adm-mat}:
\begin{equation}\label{adm-sp}
\mathbf{Y}^\text{SP} := \left(
\begin{array}{c|c|c}
\mathbf{Y}_d & -\mathbf{Y}_d\hspace{3mm} & \mathbf{0}\hspace{1mm} \\
\hline
-\mathbf{Y}_d\hspace{3mm} & \mathbf{Y}_0^{gg} + \mathbf{Y}_d & \hspace{2mm}\mathbf{Y}_0^{g\ell} \\
\hline
\mathbf{0}\hspace{1mm} & \mathbf{Y}_0^{\ell g} & \hspace{2mm}\mathbf{Y}_0^{\ell\ell}
\end{array}
\right).
\end{equation}
The equation for each generator is also of the same form but with $H_i > 0$, and is in fact exactly Eq.~\eqref{swing-eqn-cls} because the only link from the generator's internal node is to its terminal node through a purely imaginary admittance, $Y^\text{SP}_{i,i+n_g} = -(j x'_{d,i})^{-1}$, $i=1,\ldots,n_g$.
Putting together, the SP model is given by
\begin{equation}\label{eqn:sp-model-gen}
\begin{split}
&\frac{2H_i}{\omega_R}\ddot\delta + \frac{D_i}{\omega_R}\dot\delta = A^\text{SP}_i - K^\text{SP}_{i,i+n_g} \sin(\delta_i - \delta_{i+n_g}),
\quad i = 1,\ldots,n_g,\\
&A^\text{SP}_i := P_{g,i}^*,
\quad K^\text{SP}_{i,i+n_g} := |E^*_i V_i^*/x'_{d,i}|,
\end{split}
\end{equation}
for generator internal nodes and 
\begin{equation}\label{eqn:sp-model-load}
\begin{split}
&\frac{D_i}{\omega_R}\dot\delta_i 
= A^\text{SP}_i - \sum_{j=n_g+1,j\neq i}^N K^\text{SP}_{ij} \sin(\delta_i - \delta_j - \gamma^\text{SP}_{ij}),
\quad i = n_g + 1,\ldots,N,\\
&A^\text{SP}_i := - P_{\ell,i'}^* - |V_{i'}^*|^2 G^\text{SP}_{ii},
\quad K^\text{SP}_{ij} := |V_{i'}^* V_{j'}^* Y^\text{SP}_{ij}|, 
\quad i' := i - n_g, \quad j' := j - n_g,\\
&\gamma^\text{SP}_{ij} := \alpha^\text{SP}_{ij} - \frac{\pi}{2},
\quad Y^\text{SP}_{ij} = |Y^\text{SP}_{ij}| \exp(j\alpha^\text{SP}_{ij})
\end{split}
\end{equation}
for load nodes (including the generator terminal nodes), where $G^\text{SP}_{ii}$ denotes the real part of $Y^\text{SP}_{ii}$.
Note that these equations have the same form as Eq.\,\eqref{model}, and in this case the number of phase oscillators is $N = n_g + n$, since each of the generator internal and terminal nodes, as well as load nodes, is represented by a single variable $\delta_i$.
In contrast to the EN model, the SP model has the advantage that the physical structure of the transmission network represented by $\mathbf{Y}_0$ is preserved as part of the admittance matrix $\mathbf{Y}^\text{SP}$ and reflected in the coupling constants $K^\text{SP}_{ij}$.
This, however, comes at the expense of additional complexity associated with the larger set of equations (real power grids can have many more nodes than the number of generators) and  increased uncertainty associated with the additional parameters $D_i$ that must be estimated.
Note that in the SP model the transient reactances $x'_{d,i}$ are part of the preserved network structure.
Since $x'_{d,i}$ are not negligible in general (as we discussed in Sec.\,\ref{sec:eff-net} for the EN model), the corresponding links in the preserved network structure cannot be ignored, even though they do not represent any physical connections.
The inherent frequencies $\omega^*_i = \omega_R(1 + A^\text{SP}_i/D_i)$ are well defined in this model for both the internal and terminal nodes of the generators as well as for the load nodes.

If all transmission lines are assumed to be lossless and inductive (which is well-justified), the admittance matrix $\mathbf{Y}^\text{SP}$ is purely imaginary with positive imaginary component, which implies that $\gamma^\text{SP}_{ij} = 0$ for all $i \neq j$.
Under this assumption, the SP model has been used\cite{Mallada:2011lr} to design incremental adjustments to the active power output of generators or transmission line reactances to improve the linear stability of synchronous states.
The sufficient synchronization condition established in Ref.\,\onlinecite{Dorfler:2013ve} and mentioned in Sec.\,\ref{sec:eff-net} for the EN model (also assuming lossless transmission lines) has been shown in the same publication to be applicable also to the SP model.
Variations of the SP model that represent droop inverters as first-order oscillators\cite{E:2014fk} and incorporate stochastic deviations of frequencies\cite{Odun-Ayo:2012qy} have been used to study synchronization stability in networks with renewable energy sources.

\subsection{The synchronous motor model}
\label{sec:sync-motor}

Another way to model the dynamics of the loads while preserving the physical network structure is to use synchronous motors to represent the loads.
A synchronous motor is essentially the same type of machine as a generator, except that the flow of power is in the opposite direction, converting electrical power into mechanical power, and thus can be modeled by the same swing equation, Eq.~\eqref{swing-eqn} with $P_m<0$ and $P_e < 0$.
In addition to the $n_g$ internal nodes we defined for the generators in the EN model, in this case one defines $n_\ell$ internal nodes for the synchronous motors representing the loads, which makes the total number of nodes $2n$.
Since the motors are modeled in exactly the same way as the generators, this network model is mathematically equivalent to the EN model in which all $n$ nodes of the physical network has ``generators,'' $n_\ell$ of which have negative mechanical power $P_{m,i} < 0$.
The matrix $\mathbf{Y}'_0$ will be replaced by 
\begin{equation}\label{adm-mat-sm}
\mathbf{Y}''_0 := \left(
\begin{array}{c|c}
\mathbf{Y}'_d & -\mathbf{Y}'_d\hspace{3mm}\\
\hline
-\mathbf{Y}'_d\hspace{3mm} & \mathbf{Y}_0 + \mathbf{Y}'_d\\
\end{array}
\right),
\end{equation}
in this case, where $\mathbf{Y}'_d$ is the $n \times n$ diagonal matrix whose diagonal elements are the admittances $(j x'_{d,1})^{-1},\ldots, (j x'_{d,n})^{-1}$ of the transient reactances of both the generators and motors.
Kron reduction is necessary to eliminate the terminal nodes of all generators and motors, from which we obtain an $n \times n$ matrix $\mathbf{Y}^\text{SM}$ and a system of $n$ coupled phase oscillators in the same form as Eq.\,\eqref{model}.
Note that, for a given system, the matrix $\mathbf{Y}^\text{SM}$ for this model is in general different from that for the EN model.
The generator/motor transient reactances are typically non-negligible, and the network represented by $\mathbf{Y}^\text{SM}$ can be shown to have the topology of a complete graph when $x'_{d,i} > 0$ for all generators and motors, applying the same argument we used for the EN model.
The equations of motion for the SM model read
\begin{equation}
\begin{split}
&\frac{2H_i}{\omega_R}\ddot\delta_i + \frac{D_i}{\omega_R}\dot\delta_i 
= A^\text{SM}_i - \sum_{j=1,j\neq i}^{n} K^\text{SM}_{ij} \sin(\delta_i - \delta_j - \gamma^\text{SM}_{ij}),
\quad i = 1,\ldots,n,\\
&K^\text{SM}_{ij} := |E^*_i E^*_j Y^\text{SM}_{ij}|,\quad
\gamma^\text{SM}_{ij} := \alpha^\text{SM}_{ij} - \frac{\pi}{2}, \quad
Y^\text{SM}_{ij} = |Y^\text{SM}_{ij}| \exp(j\alpha^\text{SM}_{ij})
\end{split}
\end{equation}
where here $A^\text{SM}_i := P_{g,i}^* - |E^*_i|^2 G^\text{SM}_{ii}$ for generators ($i=1,\ldots,n_g$) and $A^\text{SM}_i := -P_{\ell,i}^* - |E^*_i|^2 G^\text{SM}_{ii}$ for loads/motors ($i=n_g + 1,\ldots,n$).
The internal node voltage magnitude $|E^*_i|$ for the generators and motors is determined in the same way as in the EN model.
While the representation of the loads as synchronous motors allows for a simplified description of the system with second-order differential equations for both generators and loads (in contrast to the SP model, which has both first- and second-order equations), such representation is typically not adopted in the power system engineering literature.
Most of motor loads in the US are indeed induction motors, not synchronous motors\cite{price1995standard}.

As in the case of the SP model, the well-justified assumption of lossless and inductive transmission lines implies that $\gamma^\text{SM}_{ij} = 0$ for all $i \neq j$.
This assumption is adopted in most of the previous studies that use the SM model, including the one that originally proposed the model\cite{springerlink:10.1140/epjb/e2008-00098-8}.
The SM model has been used to show that decentralization of power generation makes synchronization more sensitive to temporary increase in power demand, while simultaneously making it more robust against removal of single transmission lines\cite{PhysRevLett.109.064101}.
While transmission line removals naturally destabilize power-grid synchronization, 
it has also been shown using the same model that adding transmission lines can sometimes have the same effect\cite{Witthaut:2012wd}.
The impact of decentralization has also been shown\cite{Rohden2014ijve} to increase the order parameter that measures the degree of synchronization for a range of different random network topologies.
In conjunction with the network topology of the European power grids, the SM model has been used to study the minimum coupling strength required for synchronization\cite{Lozano:2012qy}.
Reference~\onlinecite{PhysRevE.80.066120} studies synchronization in Kuramoto oscillator networks with bipolar distribution of inherent frequencies as a model of power-grid networks (equivalent to assuming $H_i=0$ in the SM model), focusing on the effect of correlation between the inherent frequencies of two nodes that are connected. 
All of the above studies make additional simplifying assumptions in order to focus on the role of the network topology in power-grid synchronization: $H_i = H$, $D_i = D$, $P^*_{g,i} = P^*_g$, $x'_{d,i} = 0$, $|V^*_i| = 1$ p.u.\ for all $i$, and all transmission lines have identical parameters.
These assumptions together imply $|E^*_i| = 1$ p.u.\ for all $i$, $A^\text{SM}_i = P_{g,i}^*$ for generators, $A^\text{SM}_i = - P_{\ell,i}^*$ for motors, $K^\text{SM}_{ij} = K$ for all pairs of nodes $i$ and $j$ that are connected (and $K^\text{SM}_{ij} = 0$ otherwise), and $\gamma^\text{SM}_{ij} = 0$.
This corresponds to a version of the SM model for a homogeneous power grid in which all generators, motors, and transmission lines have identical parameters, but the network topology can be arbitrary.

\section{Obtaining model parameters required for simulations}
\label{sec:model-parameters}

The parameters required to run a simulation of the synchronization dynamics using Eq.\,\eqref{model} are the system reference frequency $\omega_R$, the parameters of individual generators/motors, $H_i$, $D_i$, and $x'_{d,i}$, as well as the constants for the network dynamics equation, $A_i$, $K_{ij}$, and $\gamma_{ij}$.
The parameters $\omega_R$, $H_i$, $D_i$, and $x'_{d,i}$ must be given, while $A_i$, $K_{ij}$, and $\gamma_{ij}$ are computed by solving the power flow equations\,\eqref{pf-eqn-p} and \eqref{pf-eqn-q} and performing Kron reduction.
As described in Section~\ref{sec:steady-state}, a power flow solution is usually calculated given the parameters $P^*_{g,i}$ and $|V^*_i|$ at each non-reference generator, $P^*_{\ell,i}$ and $Q^*_{\ell,i}$ at each load node, and $|V^*_i|$ at the reference generator, as well as the admittance matrix $\mathbf{Y}_0$ representing the physical network structure.
The type of power system dataset most common in engineering is suitable for this calculation and thus contains the required parameters at each node, as well as sufficient information to compute $\mathbf{Y}_0$.
Given such a dataset, standard power systems software, such as PST~\cite{chow1992toolbox} and MATPOWER~\cite{matpower}, can be used to compute a power flow solution.
The solution can then be used to build the matrix $\mathbf{Y}'_0$ incorporating the generators' $x'_{d,i}$, which can in turn be used to compute the constants $A^\text{SP}_i$, $K^\text{SP}_{ij}$, and $\gamma^\text{SP}_{ij}$ for the SP model.
From $\mathbf{Y}'_0$, $P^*_{\ell,i}$, and $Q^*_{\ell,i}$, we build $\mathbf{Y}^\text{EN}$ and then compute $A^\text{EN}_i$, $K^\text{EN}_{ij}$, and $\gamma^\text{EN}_{ij}$ for the EN model.
For the SM model, we combine the motors' $x'_{d,i}$ with the matrix $\mathbf{Y}_0$ to compute $\mathbf{Y}^\text{SM}$, which allows for the computation of $A^\text{SM}_i$, $K^\text{SM}_{ij}$, and $\gamma^\text{SM}_{ij}$.
The process of computing the model parameters required for simulations, as well as the dependency relations between all parameters involved, is illustrated in Fig.~\ref{fig:model-parameters}.
We have implemented this process using MATPOWER, and our open-source software is freely downloadable from \url{https://sourceforge.net/projects/pg-sync-models}.
The parameter computation and simulation of the dynamics can be performed by PST~\cite{chow1992toolbox} for the EN model, but not for the SP or SM models.

%:fig_model_parameters
\begin{figure*}
\begin{center}
\includegraphics{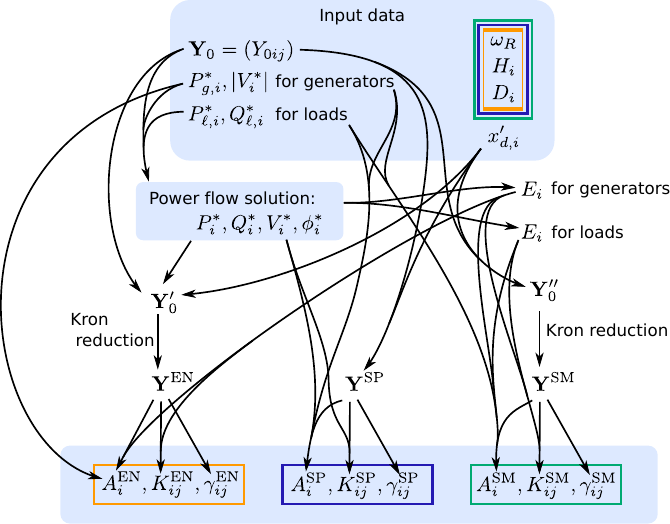}
\end{center}
\caption{\label{fig:model-parameters}
Computation of the model parameters ($A_i$, $K_{ij}$, and $\gamma_{ij}$) given the input data for simulating network synchronization dynamics and the dependency relations among the input data and the computed parameters.
The parameters required for the EN, SP, and SM models are enclosed by orange, blue, and green boxes, respectively.}
\end{figure*}

The dynamical parameters for generators and motors, $H_i$, $D_i$, and $x'_{d,i}$, are often not available.
For unavailable values of $x'_{d,i}$ and $H_i$, we suggest estimating them using the strong correlation between each of these parameters and the active power output $P^*_{g,i}$ of generator $i$, which was observed in Ref.\,\onlinecite{Motter:2013fk} for the first three systems listed in Table~\ref{table:systems}.
(The values for these three systems are available from Refs.\,\onlinecite{Anderson:1986fk,pai1989energy,Vittal:1992lr}.)
The parameters are estimated as
$x'_{d,i} = 92.8 (P^*_{g,i})^{-1.3}$  and $ H_i = 0.04 P^*_{g,i}$, while imposing a maximum value of one for $x'_{d,i}$ and a minimum value of $0.1$ for $H_i$, where $P^*_{g,i}$ is measured in megawatts, $x'_{d,i}$ is in p.u., and $H_i$ is in seconds.
Our method of estimating $x'_{d,i}$ is a refinement of a standard approach based on the observation that real values of $x'_{d,i}$ tend to fall within a narrow range when expressed in p.u. with respect to the rated power of the generator/motor\cite{Anderson:1986fk,Grainger:1994yo}. This observation, combined with the assumption that the rated power (which is unavailable in many datasets) is proportional to $P^*_{g,i}$, implies that $x'_{d,i}$ is inversely proportional to $P^*_{g,i}$ when $x'_{d,i}$ is expressed in p.u.\ with respect to a reference power common to the entire network. In our previous work\cite{Motter:2013fk} we found that the available data for the first three systems in Table 1 follows our formula more closely than the standard approach.
For $D_i$, one may use $D_i = (D_{m,i} + D_{e,i} + 1/R_i)\omega_R^2/P_R = 50$~p.u., a value that results from assuming that mechanical friction and the electrical effect of damper windings are negligible ($D_{m,i} = D_{e,i} = 0$) compared to the effect of regulation by governors with $R_i = 0.02 \,\omega_R^2/P_R = 0.02$~p.u.

%:Table
\begin{table*}
\caption{Structural and dynamical properties of the systems considered.}
\label{table:systems}
\begin{tabular*}{\hsize}{@{\extracolsep{\fill}}lrrrrrrr}
\hline\hline
& & & & & \multicolumn{3}{c}{Number of oscillators}\\
\cline{6-8}
System & Nodes & Links & Generators & Loads & \hspace{3mm}EN & \hspace{7mm}SP & SM \\
\hline
3-generator test system (Ref.\,\onlinecite{Anderson:1986fk}) & 9 & 9 & 3 & 6 & 3 & 12 & 9\\
10-generator test system (Ref.\,\onlinecite{pai1989energy}) & 39 & 46 & 10 & 29 & 10 & 49 & 39\\
50-generator test system (Ref.\,\onlinecite{Vittal:1992lr}) & 145 & 422 & 50 & 95 & 50 & 195 & 145\\
Guatemala power grid\footnote[1]{Data provided by F.\ Milano (University of Castilla -- La Mancha).} & 370 & 392 & 94 & 276 & 94 & 464 & 370\\
Northern Italy power grid\footnotemark[1]  & 678 & 822 & 170 & 508 & 170 & 695 & 678\\
Poland power grid (Ref.\,\onlinecite{matpower}) & 2383 & 2886 & 327 & 2056 & 327 & 2710 & 2383\\
\hline\hline
\end{tabular*}
\end{table*}

\section{Heterogeneity of model parameters}
\label{sec:hetero}

Given the input data for the system, the constants $A_i$, $K_{ij}$, and $\gamma_{ij}$ for the network dynamics governed by Eq.\,\eqref{model} can generally be distributed heterogeneously across the network.
Since $A_i$ are related to the inherent frequencies $\omega^*_i$ through the same formula $\omega^*_i = \omega_R(1 + A_i/D_i)$ for all three models, the $A_i$-heterogeneity quantifies the heterogeneity of the intrinsic properties of the generator dynamics.
While the heterogeneity in the generator parameters is naturally a source of heterogeneity in $A_i$, it is actually possible to have heterogeneous $A_i$ even in the complete absence of heterogeneity in the generator parameters $H_i$, $D_i$, $x'_{d,i}$, $P^*_{g,i}$, and $|V^*_i|$.
The first two parameters actually do not enter into the calculation of $A_i$ (as described in the previous section) and thus do not affect the $A_i$-heterogeneity.
The other three affect the values of $A_i$, but do not fully determine $A_i$, since the calculation also involves the admittance matrix $\mathbf{Y}_0$ as well as $P^*_{\ell,i}$ and $Q^*_{\ell,i}$ for the loads.
In the following subsections, we first show that $A_i$-heterogeneity can indeed arise without heterogeneity of generator parameters using a small network example, and generalize it to larger networks.
We then discuss the $A_i$-heterogeneity as well as the coupling matrices $K_{ij}$ and $\gamma_{ij}$ for examples of realistic systems.

\subsection{System with two generators and one load}\label{sec:two-ten-one-load}

We consider the simplest case with which the heterogeneity of generator parameters can be discussed: a system with two generators, each connected to a single load by a transmission line, as shown in Fig.~\ref{fig:modeling}.
For this system we have $n_g = 2$, $n_\ell = 1$, $n = n_g + n_\ell = 3$, and $N = 2n_g + n_\ell = 5$.
Assuming lossless and inductive transmission lines (i.e., $r_{13} = r_{23} = 0$, $x_{13} >0$, and $x_{23} > 0$), the power flow equations for this network read
\begin{align}
P^*_{g,1} &= \left\vert\frac{V^*_1 V_3}{x_{13}}\right\vert\sin(\phi_1 - \phi_3),\label{eqn:pf2-gen-1}\\
P^*_{g,2} &= \left\vert\frac{V^*_2 V_3}{x_{23}}\right\vert\sin(\phi_2 - \phi_3),\label{eqn:pf2-gen-2}\\
- P^*_{\ell,1} &= - P^*_{g,1} - P^*_{g,2},\\ 
Q_1 &= - |V^*_1|^2 B_{011} 
- \left\vert\frac{V^*_1 V_3}{x_{13}}\right\vert\cos(\phi_1 - \phi_3),\\
Q_2 &= - |V^*_2|^2 B_{022} 
- \left\vert\frac{V^*_2 V_3}{x_{23}}\right\vert\cos(\phi_2 - \phi_3),\\
-Q^*_{\ell,3} &= - |V_3|^2 B_{033} 
- \left\vert\frac{V_3 V^*_1}{x_{13}}\right\vert\cos(\phi_3 - \phi_1)
- \left\vert\frac{V_3 V^*_2}{x_{23}}\right\vert\cos(\phi_3 - \phi_2),
\end{align}
where $B_{011} = \text{Im}(Y_{011}) = \frac{b_{13}}{2} - \frac{1}{x_{13}} > 0$, $B_{022} = \frac{b_{23}}{2} - \frac{1}{x_{23}} > 0$, and $B_{033} = B_{011} + B_{022}$.
Setting $\phi_1 = 0$ and using it as the reference angle, we solve this set of five equations for five unknowns, $\phi_2$, $\phi_3$, $Q_1$, $Q_2$, $|V_3|$ to obtain a power flow solution.
(The third equation does not involve unknown variables and simply imposes a constraint on the parameters.)
The equations of motion for the three network dynamics models are:

\medskip\noindent EN model:
\begin{equation}\label{eqn:two-gen-EN}
\begin{split}
&\frac{2H_1}{\omega_R}\ddot\delta_1 + \frac{D_1}{\omega_R}\dot\delta_1 
= A^\text{EN}_1 - K^\text{EN} \sin(\delta_1 - \delta_2 - \gamma^\text{EN}),\\
&\frac{2H_2}{\omega_R}\ddot\delta_2 + \frac{D_2}{\omega_R}\dot\delta_2 
= A^\text{EN}_2 - K^\text{EN} \sin(\delta_2 - \delta_1 - \gamma^\text{EN}),\\
\end{split}
\end{equation}
where
\begin{align}
&A^\text{EN}_1 = P_{g,1}^* - |E^*_1|^2 G^\text{EN}_{11},
\quad A^\text{EN}_2 = P_{g,2}^* - |E^*_2|^2 G^\text{EN}_{22}\nonumber\\
&K^\text{EN} = |E^*_1 E^*_2 Y_{12}|,\quad
\gamma^\text{EN} = \alpha^\text{EN}_{12} - \frac{\pi}{2},
\quad Y^\text{EN}_{12} = |Y^\text{EN}_{12}|\exp(j\alpha^\text{EN}_{12}).\nonumber
\end{align}
\newpage\noindent SP model:
\begin{equation}\label{eqn:two-gen-SP}
\begin{split}
\frac{2H_1}{\omega_R}\ddot\delta_1 + \frac{D_1}{\omega_R}\dot\delta_1 
&= A^\text{SP}_1 - K^\text{SP}_{13} \sin(\delta_1 - \delta_3),\\
\frac{2H_2}{\omega_R}\ddot\delta_2 + \frac{D_2}{\omega_R}\dot\delta_2 
&= A^\text{SP}_2 - K^\text{SP}_{24} \sin(\delta_2 - \delta_4),\\
\frac{D_3}{\omega_R}\dot\delta_3
&= A^\text{SP}_3 - K^\text{SP}_{31} \sin(\delta_3 - \delta_1) - K^\text{SP}_{35} \sin(\delta_3 - \delta_5),\\
\frac{D_4}{\omega_R}\dot\delta_4
&= A^\text{SP}_4 - K^\text{SP}_{42} \sin(\delta_4 - \delta_2) - K^\text{SP}_{45} \sin(\delta_4 - \delta_5),\\
\frac{D_5}{\omega_R}\dot\delta_5
&= A^\text{SP}_5 - K^\text{SP}_{53} \sin(\delta_5 - \delta_3) - K^\text{SP}_{54} \sin(\delta_5 - \delta_4),
\end{split}
\end{equation}
where
\begin{align}
&\hspace{5mm} A^\text{SP}_1 = P_{g,1}^*, \quad A^\text{SP}_2 = P_{g,2}^*,\nonumber\\
&\hspace{5mm} A^\text{SP}_3 := - |V_1^*|^2 G'_{033}, \quad A^\text{SP}_4 := - |V_2^*|^2 G'_{044},
\quad A^\text{SP}_5 := - P_{\ell,3}^* - |V_3^*|^2 G'_{055},\nonumber\\
&\hspace{5mm} K^\text{SP}_{13} = K^\text{SP}_{31} = |E_1 V_1^*/x'_{d,1}|,
\quad K^\text{SP}_{24} = K^\text{SP}_{42} = |E_2 V_2^*/x'_{d,2}|,\nonumber\\
&\hspace{5mm} K^\text{SP}_{35} = K^\text{SP}_{53} = |V_1^* V_3^* G'_{035}|,
\quad K^\text{SP}_{45} = K^\text{SP}_{54} = |V_2^* V_3^* G'_{045}|.\nonumber
\end{align}
\medskip\noindent SM model:
\begin{equation}\label{eqn:two-gen-SM}
\begin{split}
&\frac{2H_1}{\omega_R}\ddot\delta_1 + \frac{D_1}{\omega_R}\dot\delta_1 
= A^\text{SM}_1 - K^\text{SM}_{12} \sin(\delta_1 - \delta_2) - K^\text{SM}_{13} \sin(\delta_1 - \delta_3),\\
&\frac{2H_2}{\omega_R}\ddot\delta_2 + \frac{D_2}{\omega_R}\dot\delta_2
= A^\text{SM}_2 - K^\text{SM}_{21} \sin(\delta_2 - \delta_1) - K^\text{SM}_{23} \sin(\delta_2 - \delta_3),\\
&\frac{2H_3}{\omega_R}\ddot\delta_3 + \frac{D_3}{\omega_R}\dot\delta_3 
= A^\text{SM}_3 - K^\text{SM}_{31} \sin(\delta_3 - \delta_1) - K^\text{SM}_{32} \sin(\delta_3 - \delta_2),
\end{split}
\end{equation}where
\begin{align}
&\hspace{5mm} A^\text{SM}_1 = P_{g,1}^* - |E^*_1|^2 G^\text{SM}_{11},
\quad A^\text{SM}_2 = P_{g,2}^* - |E^*_2|^2 G^\text{SM}_{22},
\quad A^\text{SM}_3 = - P_{\ell,3}^* - |E^*_3|^2 G^\text{SM}_{33},\nonumber\\
&\hspace{5mm} K^\text{SM}_{ij} := |E^*_i E^*_j Y^\text{SM}_{ij}|, \quad i \neq j, \,\, i,j = 1,2,3.\nonumber
\end{align}
Note that the phase shifts $\gamma^\text{SP}_{ij}$ and $\gamma^\text{SM}_{ij}$ do not appear in the equations, since the assumption of lossless and inductive transmission lines implies that they are both zero, as discussed in Sec.\,\ref{sec:struct-preserving} and Sec.\,\ref{sec:sync-motor}.

Assuming that $P^*_{g,1}$, $P^*_{g,2}$, $|V^*_1|$, and $|V^*_2|$ are all nonzero, Eqs.\,\eqref{eqn:pf2-gen-1} and \eqref{eqn:pf2-gen-2} imply
\begin{equation}
\frac{P^*_{g,1}|V^*_2 x_{13}|}{P^*_{g,2}|V^*_1 x_{23}|}
= \frac{\sin(\phi_1 - \phi_3)}{\sin(\phi_2 - \phi_3)}
\end{equation}
(regardless of the choice of the network dynamics model).
Thus, even if the two generators' output power and voltage magnitudes are identical, the steady-state phase angle differences (with respect to the phase of the load node) can be different if the parameters of the transmission lines are different.
This has an interesting implication for the EN model: the oscillators can have different parameters (i.e., $A^\text{EN}_1 \neq A^\text{EN}_2$, and therefore different inherent frequencies) even when the generator they represent have identical parameters (i.e., $P^*_{g,1} = P^*_{g,2}$, $V^*_1 = V^*_2$, and $x'_{d,1} = x'_{d,2}$) if the transmission lines have different parameters. 
This is because we have $A^\text{EN}_i = P^*_i - |E^*_i|^2 G^\text{EN}_{ii}$, the constant $|E^*_i|$ depends on the steady-state phase angles, and $G^\text{EN}_{ii}$ depends on the transmission line parameters.
The effect is illustrated in Fig.~\ref{fig:modeling}C using the following concrete values of the system parameters: $x_{13} = 0.04$~p.u., $x_{23} = 0.047$~p.u., $b_{13} = 0.082$~p.u., $b_{23} = 0.098$~p.u., $P^*_{g,1} = P^*_{g,2} = 1$~p.u.\ $= 100$~MW, $P^*_{\ell,3} = 2$~p.u.~$=200$~MW, $Q^*_{\ell,3} = 1$~p.u.~$=100$~MVAR, $V^*_1 = V^*_2 = 1$~p.u.~$= 100$~kV, $x'_{d,1} = x'_{d,2} = 0.1$~p.u.\ (with all p.u.\ quantities with respect to the 100 MVA system base).
If we lift the assumption of lossless transmission lines (purely imaginary impedances), the effect we have demonstrated for the EN model can also be observed for the SM model.
Indeed, if we set $r_{13} = 0.002$~p.u.\ and $r_{23} = 0.007$~p.u., while adjusting $P^* := P^*_1 = P^*_2$ so that the power flow equation is satisfied and keeping all the other parameters the same, we obtain $A^\text{SM}_1 = 0.9217$ and $A^\text{SM}_2 = 0.8109$ (with $P^* = 1.0055$~p.u.) for the SM model.
In contrast, the same effect cannot be observed for the SP model, since we have $A^\text{SP}_i = P_{g,i}^*$ for all generators regardless of the network structure.

The effect of network asymmetry demonstrated here for this small example system when using the EN and SM models suggests generalization to larger systems, as well as to other forms of network asymmetry, such as heterogeneities in the distribution of power demand, generator locations, and the connectivity of generators/loads in terms of the number of transmission lines.
This will be discussed in the next section.

\subsection{Larger power-grid networks}

Consider a system with identical generator parameters: $P^*_{g,i} = P^*_g$, $|V^*_i| = |V^*|$, and $x'_{d,i} = x'_d$ for all $i = 1,\ldots,n_g$.
Consider the EN model.
The constants $A^\text{EN}_i = P_g^* - |E^*_i|^2 G^\text{EN}_{ii}$ are identical if and only if $|E^*_i|^2 G^\text{EN}_{ii}$ are identical.
From Eq.\,\eqref{eqn-E}, we see that requiring identical $|E^*_i|$ constrains all $Q^*_i$.
Also, requiring $G^\text{EN}_{ii}$ to be identical is equivalent to a global constraint through Eq.\,\eqref{eq:eff-adm}.
We thus see that $A^\text{EN}_i$ will not be identical for a power flow solution in a generic setting.
Nongeneric situations can arise if there is a symmetry in the system, such as the system with the star configuration in which all generators are connected to a single load through  identical transmission lines.
Even with this star configuration, any heterogeneity in the parameters of the transmission lines will lead to heterogeneity in $\phi^*_i$ and $Q^*_i$ at the generator terminal nodes, and thus in $|E^*_i|$ and $A^\text{EN}_i$.
A similar argument can be made for the SM model.
We therefore conclude that a generic system has heterogeneous $A_i$ under both the EN model and the SM model.

We now study the heterogeneity of the model parameters for the set of realistic systems listed in Table~\ref{table:systems}, which are used in Ref.\,\onlinecite{Motter:2013fk} to study the stability of synchronous states against small perturbations.
The data required for power flow calculations for these systems were obtained as follows.
For the 10-generator system, known as the New England test system, the parameters were taken from Ref.\ \onlinecite{pai1989energy}. 
For the 3- and 50-generator systems, the parameters were taken from 
Refs.\ \onlinecite{Anderson:1986fk} 
and \onlinecite{Vittal:1992lr}, 
respectively.  The data for the Guatemala and Northern Italy systems were provided by F.\ Milano (University of Castilla -- La Mancha), 
and the data for the Poland system were provided as part of the MATPOWER software package\cite{matpower}.  
The required dynamic data that are not available from the respective sources cited above were computed as described in Section~\ref{sec:model-parameters}. 

%:fig_compare_nat_freq
\begin{figure*}
\begin{center}
\includegraphics[width=\textwidth]{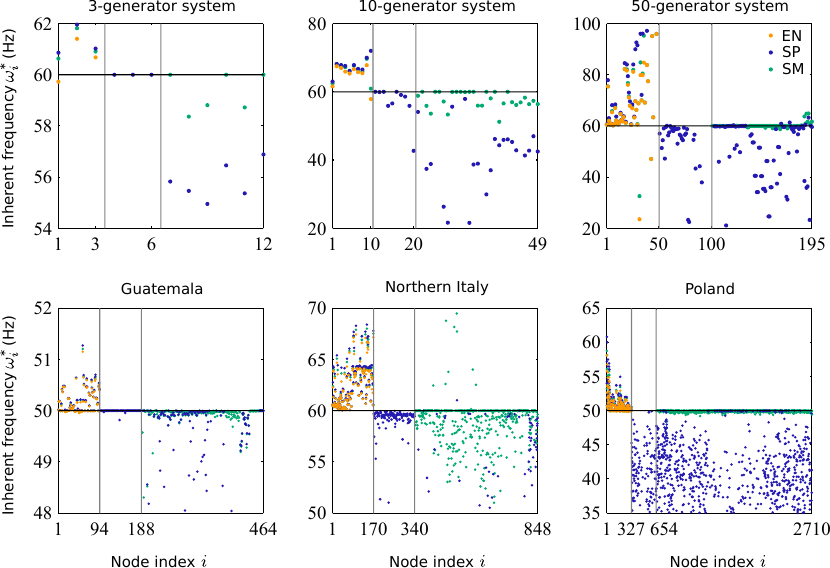}
\end{center}
\vspace{-5mm}
\caption{\label{fig-compare-nat-freq}
The inherent frequencies $\omega_i^*$ of the nodes in the systems listed in Table~\ref{table:systems} under the EN (orange), SP (blue), and SM (green) models.
The nodes are indexed so that $i=1,\ldots,n_g$ correspond to the generator internal nodes, the $i=n_g+1,\ldots,2n_g$ to the generator terminal nodes, and $i=2n_g+1,\ldots,N (= 2n_g + n_\ell)$ to the load nodes.
The three types of nodes are separated by the two gray vertical lines at $i = n_g$ and $i = 2n_g$. 
The plots show $\omega_i^*$ of the generator internal nodes for all three models, the generator terminal nodes only for the SP model, and the load nodes only for the SP and SM models.
The black horizontal lines indicate each system's reference frequency.
We note that the inherent frequency of some generators and loads are outside the range of the plots.}
\end{figure*}

Figure~\ref{fig-compare-nat-freq} shows the computed inherent frequencies $\omega^*_i = \omega_R(1 + A_i/D_i)$ under the three models for each of the systems in Table~\ref{table:systems}.
Note that those realistic systems have heterogeneous generator parameters, which is a source of heterogeneity in $A_i$ in addition to the effect described above that results from the heterogeneity and asymmetry of the physical network structure.
We see that the inherent frequencies can be significantly different from the system reference frequency $\omega_R$, with mostly higher frequencies for the generators and lower frequencies for the load nodes (except for the EN model, in which the load nodes are eliminated and thus have no well-defined inherent frequencies).
This is consistent with the general picture that the voltage phase angles of the generators tend to rotate at higher frequencies and pull the frequency of the rest of the grid upward, while the load phase angles tend to rotate at lower frequencies and pull the rest of the grid downward.
The variation of the inherent frequencies among generators and load nodes can be extremely large.
Indeed, we find a generator with $\omega^*_i > 7 \times 10^2$~Hz for the 50-generator system, and a load node with $\omega^*_i < -1.1 \times 10^3$~Hz (corresponding to a rotation in the opposite direction with frequency $> 1.1 \times 10^3$~Hz) for the Northern Italy system.
As noted in Sec.\,\ref{sec:intro}, frequencies far from $\omega_R$ are not likely to be observed in reality, but these values of inherent frequency do reflect the inherent dynamical property of the oscillators representing generators or loads.
The observed large variation in inherent frequency across the network may indicate a high level of frustration in the system in the sense that oscillators with vastly different frequencies are forced to rotate at a common frequency by the coupling.
Comparing across different network dynamics models, we see similarity in the pattern of generator-to-generator variation, as well as noticeable differences in the inherent frequency of the same generator or load under different models.
The difference across models does not always depend monotonically on the size of the system, as evident from comparing the generator frequencies in the 10-generator system, 50-generator system, and Guatemala power grid.

%:fig_compare_K
\begin{figure*}
\begin{center}
\includegraphics[width=\textwidth]{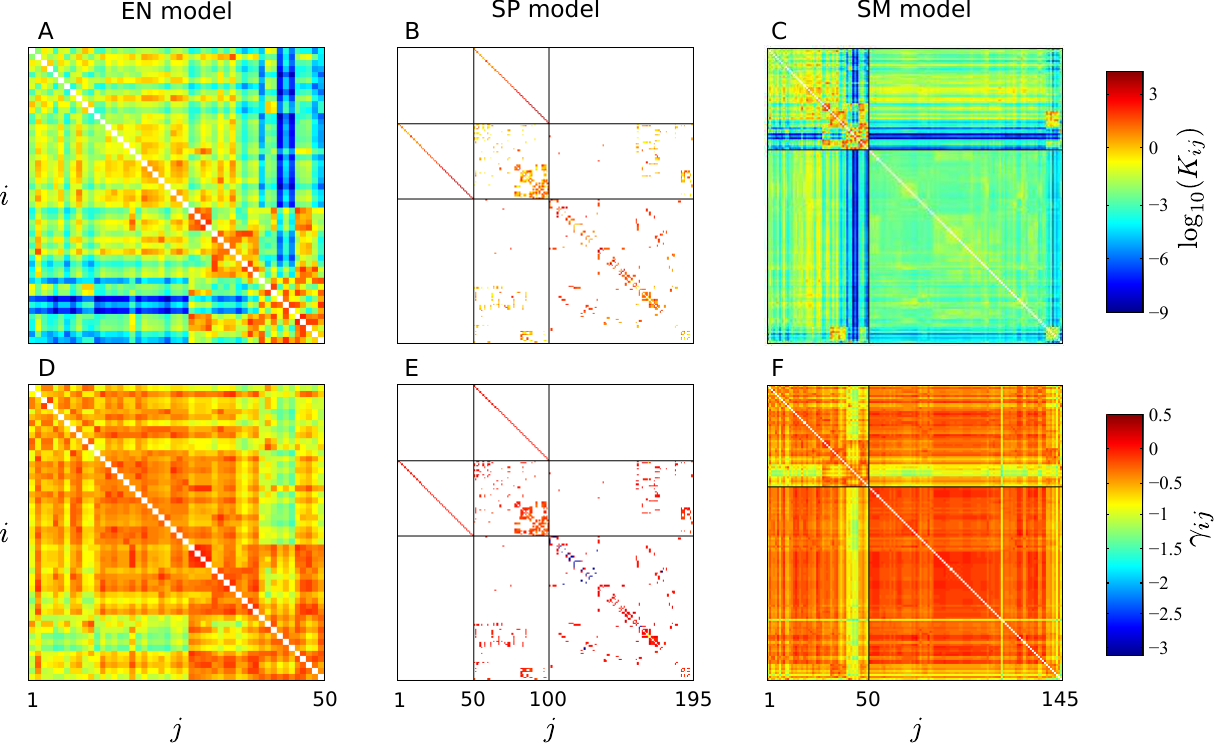}
\end{center}
\caption{\label{fig-compare-K}
(A)--(C) Coupling matrix $\mathbf{K} = (K_{ij})$ for the three different models of the 50-generator system (with $n_g = 50$, $n_\ell = 95$, $n = 145$, $N = 195$).  (D)--(F) The corresponding phase shift matrix $\boldsymbol{\Gamma} = (\gamma_{ij})$.
In each row, the common color scale used for all three matrices corresponding to the three models is shown on the right.
The matrix elements are colored white if they do not affect the network dynamics (those elements for which $i = j$ or $K_{ij} = 0$). }
\end{figure*}

Figure~\ref{fig-compare-K} shows a visualization of the coupling matrix $\mathbf{K} = (K_{ij})$ and the phase shift matrix $\mathbf{\Gamma} = (\gamma_{ij})$ for the 50-generator system. 
Recall first that the size of the coupling and phase shift matrices is different for different models: $n_g \times n_g$ for the EN model, $N \times N$ for the SP model (where $N = 2n_g+n_\ell$), and $n \times n$ for the SM model (where $n = n_g + n_\ell$).
Despite the fundamental differences in the dimensionality and in the type of equations used in the three models, we can identify some similarity within the coupling structure. 
We see in Fig.~\ref{fig-compare-K}A that there is a non-trivial structure in the dynamical coupling among the generator internal nodes under the EN model.
Although this coupling matrix is fully-populated, the strength of coupling varies widely and spans across many orders of magnitude, from $10^{-9}$ to order one.
Notice that a similar coupling structure can be identified in the middle box in Fig.~\ref{fig-compare-K}B, corresponding to the sparse physical network connecting the generator terminal nodes in the SP model.
We can also identify a similar structure in the top left box in Fig.~\ref{fig-compare-K}C for the SM model.
Most of the $\gamma_{ij}$ values are close to zero (red in Fig.~\ref{fig-compare-K}D--F),
which indicates that the coupling between oscillators $i$ and $j$ tends to make the oscillators synchronized with a small phase angle difference.
A zero phase shift corresponds to an admittance (physical or effective, depending on the model) that has positive imaginary part and zero real part, which means that it represents a pure capacitive reactance.
There are, however, some pairs of oscillators for which $\gamma_{ij}$ is significantly different from zero (green to blue in Fig.~\ref{fig-compare-K}D--F).
Such large phase shifts correspond to resistive or inductive components.

\section{Conclusions}
\label{sec:conclusions}

We have presented first-principle derivations for three different network-level models of power-grid synchronization dynamics.
These models are all based on the same fundamental equation of motion for mechanical rotation (the swing equation) and the same electric circuit representation (the classical model) of power generators and synchronous motors.
The models are, however, clearly distinguished by the different sets of assumptions adopted for the loads.
All three models are formulated in their full generality, which helps clarify how different choices of assumptions lead to different special cases of these models that have been used in previous studies.
We have shown that even when we assume all the generators to be completely identical, the parameters of the phase oscillators representing them in the EN and SM models may be quite different.
We have also shown that the oscillator parameters are vastly heterogeneous in a selection of test systems and real power-grid datasets.

While the validity and appropriateness of the models depend on the purpose, the choice of additional assumptions, and the system under consideration, these three models lay a solid foundation for the study of synchronization dynamics in complex power-grid networks.
If the purpose is to isolate the effect of the network topology on synchronization, making simplifying assumptions (as has been done in many previous studies) to homogenize all other aspects of the system may be appropriate.
If the purpose is to investigate the interplay between heterogeneities in the parameters of the generators, loads, and transmission lines, it may be helpful to use a simple, random, and/or regular network topology.
If the purpose is to understand the network structure and the component parameters of realistic power systems, then keeping the details in a given dataset and comparing the synchronization properties with null models will provide the most useful insights.
The models we considered here can serve as bases for stability analysis against small or large perturbations, for which a variety of analytical and numerical techniques are available. 
As shown in Ref.\,\onlinecite{Motter:2013fk}, stability against small perturbations can be fully addressed using the powerful conceptual framework of master stability functions\cite{Pecora:1998zp} based on linear stability analysis and network spectral theory.
In contrast, results on stability against large perturbations tend to be more limited in scope. If the relevant admittance matrix has zero real components, the problem can be tackled by energy function-based approaches\cite{pai1989energy}, which can also be used in combination with bi-stable representation of transmission lines for the modeling of the dynamics of cascading failure propagation in power grids\cite{DeMarco:2001lr}. Other approaches include estimating the region of attraction of a synchronous state\cite{Gajduk:2014kx}, quantifying the size of the region using a recently proposed measure called the basin stability\cite{Menck:2013fk}, and analyzing nonlinear modes using Koopman operators~\cite{Mezic:2005fj,Susuki:2011qy}.
Such analyses of synchronization stability against small and large perturbations may help develop strategies for improving the existing power grids by elucidating the relation between network structure and stability, which we have recently demonstrated to be highly non-intuitive in general models of coupled oscillator networks~\cite{Nishikawa:2010fk}.

Understanding the dynamics of the models we considered here can potentially serve as a stepping stone for studying other types of instabilities relevant for power grids.
For example, studying the so-called dynamic bifurcation problem\cite{E:1991lr} for these models, in which slow variation of the parameters $A_i$, $K_{ij}$, and/or $\gamma_{ij}$ drives the system toward a bifurcation point, can provide insights into the issue of voltage instability in power grids\cite{Dobson1992kls}.
To capture the fast dynamics that follows the bifurcation, one may supplement the models we studied with additional equations that describe time-varying voltage magnitudes\cite{Chiang1990kgs,Schmietendorf:2014ys,Nagata:2014vn}. 
Such model-based studies of voltage instability complement model-free approaches for predicting the bifurcation point based on time series measurement\cite{Cotilla-Sanchez:2012fk}.
Other causes for instability include fluctuations in power demand and failure of system components, and modeling them as stochastic processes can help optimize response strategies for power grid operators~\cite{Anghel:2007uq}.
With the anticipated transition to smart grids, there will be additional fluctuations in the system from new components such as intermittent energy sources and increased use of plug-in electric vehicles\cite{Nardelli:2014uq}.
Such fluctuations will have impact on the synchronization stability of power grids, and its implications will be difficult to understand unless we understand the short-term dynamics of the models considered here.
By providing comprehensive comparative analysis of these models at the intersection of the physics of complex systems, network science, and power system engineering, we hope to contribute to a better understanding and control of the dynamics of existing power grids as well as to a better design of future power grids.

\begin{acknowledgments}
The authors thank F.~Milano for providing power-grid data and the guest editors of the focus issue, M.~Timme, L.~Kocarev, and D.~Witthaut, for organizing this timely publication.
This work was supported by a Booster Award from the Institute for Sustainability and Energy at Northwestern (ISEN) and the U.S. National Science Foundation under Grant DMS-1057128.
\end{acknowledgments}

\appendix*

\section{Single generator connected to a large power grid}
\label{sec:single-gen-example}

Consider a generator connected through a single transmission line to a large power grid.
Since the power grid is large, the dynamics of the generator does not affect the behavior of the rest of the grid much, and we thus approximate the voltage magnitude and the phase angular frequency at the other end of the line to be constant.
This approximation corresponds to having the generator connected to a sort of ``heat bath,'' which can be modeled as an artificial generator that has infinite inertia and can absorb an arbitrary amount of active and reactive power without changing its terminal voltage magnitude and phase frequency (and thus is often called an infinite bus, where the term ``bus'' is used by power system engineers to refer to a network node).
We measure the generator's phase angle $\delta$ relative to the phase of the voltage at the power-receiving end of the transmission line, which has magnitude $|V_\infty|$ and phase rotating at the reference frequency $\omega_R$.
Combining the transient reactance and the transmission line impedance $Z = R + jX$ to obtain a total admittance of $Y = [R+j(X + x'_d)]^{-1}= |Y|e^{j\alpha} = G + jB$ (which defines $\alpha$), the active power output $P_e$ from the generator at its internal node can be calculated as $P_e = |E^*|^2 G + |E^* V_\infty Y|\sin(\delta - \gamma)$, where $\gamma := \alpha - \pi/2$.
Equation~\eqref{swing-eqn} then becomes
\begin{equation}\label{single-gen}
\frac{2H}{\omega_R}\ddot\delta + \frac{D}{\omega_R}\dot\delta = A - K\sin(\delta - \gamma),
\end{equation}
where $A = P_m - |E^*|^2 G$ and $K = |E^* V_\infty Y|$.
Upon changing the variable to $\theta := \delta - \gamma$ and defining $M := 2H/\omega_R$ and $\hat{D} := D/\omega_R$, the equation becomes $M\ddot\theta + \hat{D}\dot\theta = A - K\sin\theta$, which is the equation of motion for a forced damped pendulum, where $\theta$ is the angle the pendulum arm makes with the vertical, $M$ is the mass, $\hat{D}$ is the damping coefficient, and $A$ is the forcing torque.
We thus see that a pair of stable and unstable equilibria exist if and only if $A < K$, and they are characterized by $\dot\theta^* = 0$ and $A = K\sin\theta^*$ and correspond to the states of the generator synchronized to that of the rest of the grid at the frequency $\omega_R$.

If $\hat{D}^2 > 4M\sqrt{K^2-A^2}$, the effect of damping is strong enough, and $\dot\delta$ will converge exponentially to zero (corresponding to the generator's instantaneous frequency converging to $\omega_R$) after a small perturbation is applied when the system is at the stable equilibrium.
Otherwise, the damping is too weak, and the frequency will oscillate around $\omega_R$ with decaying amplitude with a characteristic frequency determined by the parameters.
When $\hat{D}=0$, the equilibrium is neutrally stable and the (angular) frequency of oscillation around the equilibrium is $(K^2 - A^2)^{1/4}/\sqrt{M}$, which is referred to as the ``natural frequency'' of the generator in the power systems engineering literature\cite{Anderson:1986fk}.
The natural frequency is different from the inherent frequency discussed in the main text, which gives $\omega^* = \omega_R + (P_m - |E^*|^2G)/\hat{D}$ (corresponding to $\dot\delta^* = A/\hat{D}$) and corresponds to the equilibrium frequency of the generator's rotor in the absence of the coupling term (i.e., $K=0$).
The $K=0$ decoupling situation can in fact be realized briefly if a fault occurs on the transmission line near the other end and creates a short circuit to the ground.

We emphasize here that, among the parameters that determine the dynamics of Eq.~\eqref{single-gen}, the constants $|E^*|$ and $P_m$ depend on how much power is flowing over the transmission line, in contrast to all the others, which represent the property of the system itself and are independent of the power flow.
In a practical setting, the power flow over transmission lines is determined by solving the power flow equations~\eqref{pf-eqn-p} and \eqref{pf-eqn-q}, given the active power output and the voltage magnitude at the generator terminal nodes as well as the active
and reactive power consumption at the load nodes.
Given the voltage magnitude $|V^*|$ and the active
power $P_g^*$ injected by the generator at the terminal into the transmission line in the single generator example of Eq.~\eqref{single-gen}, Eqs.~\eqref{pf-eqn-p} and \eqref{pf-eqn-q} (with $n=2$) give
\begin{align}
P_g^* = \frac{|V^* V_\infty|}{|Z|}\cos(\phi + \alpha),
\quad Q_g^* = - \frac{|V^* V_\infty|}{|Z|}\sin(\phi + \alpha),
\end{align}
where $\phi$ is the voltage phase angle at the terminal relative to that at the other end of the transmission line and $Q_g^*$ is the reactive power injected into the line.
This can be used to express $Q_g^*$ in terms of the given parameters as $Q_g^* = \sqrt{|V^* V_\infty|^2/|Z|^2 - (P_g^*)^2}$. 
This, together with $P_m = P_g^*$ and Eq.~\eqref{eqn-E}, determines the dependence of the parameters of Eq.~\eqref{single-gen} on the steady-state power flow of the system, specifically as a function of $P_g^*$, $|V^*|$, and $|V_\infty|$.

%:References
\bibliography{references.bib}

\begin{thebibliography}{52}%
\makeatletter
\providecommand \@ifxundefined [1]{%
 \@ifx{#1\undefined}
}%
\providecommand \@ifnum [1]{%
 \ifnum #1\expandafter \@firstoftwo
 \else \expandafter \@secondoftwo
 \fi
}%
\providecommand \@ifx [1]{%
 \ifx #1\expandafter \@firstoftwo
 \else \expandafter \@secondoftwo
 \fi
}%
\providecommand \natexlab [1]{#1}%
\providecommand \enquote  [1]{``#1''}%
\providecommand \bibnamefont  [1]{#1}%
\providecommand \bibfnamefont [1]{#1}%
\providecommand \citenamefont [1]{#1}%
\providecommand \href@noop [0]{\@secondoftwo}%
\providecommand \href [0]{\begingroup \@sanitize@url \@href}%
\providecommand \@href[1]{\@@startlink{#1}\@@href}%
\providecommand \@@href[1]{\endgroup#1\@@endlink}%
\providecommand \@sanitize@url [0]{\catcode `\\12\catcode `\$12\catcode
  `\&12\catcode `\#12\catcode `\^12\catcode `\_12\catcode `\%12\relax}%
\providecommand \@@startlink[1]{}%
\providecommand \@@endlink[0]{}%
\providecommand \url  [0]{\begingroup\@sanitize@url \@url }%
\providecommand \@url [1]{\endgroup\@href {#1}{\urlprefix }}%
\providecommand \urlprefix  [0]{URL }%
\providecommand \Eprint [0]{\href }%
\providecommand \doibase [0]{http://dx.doi.org/}%
\providecommand \selectlanguage [0]{\@gobble}%
\providecommand \bibinfo  [0]{\@secondoftwo}%
\providecommand \bibfield  [0]{\@secondoftwo}%
\providecommand \translation [1]{[#1]}%
\providecommand \BibitemOpen [0]{}%
\providecommand \bibitemStop [0]{}%
\providecommand \bibitemNoStop [0]{.\EOS\space}%
\providecommand \EOS [0]{\spacefactor3000\relax}%
\providecommand \BibitemShut  [1]{\csname bibitem#1\endcsname}%
\let\auto@bib@innerbib\@empty
%</preamble>
\bibitem [{\citenamefont {Dorogovtsev}, \citenamefont {Goltsev},\ and\
  \citenamefont {Mendes}(2008)}]{Dorogovtsev:2008ly}%
  \BibitemOpen
  \bibfield  {author} {\bibinfo {author} {\bibfnamefont {S.~N.}\ \bibnamefont
  {Dorogovtsev}}, \bibinfo {author} {\bibfnamefont {A.~V.}\ \bibnamefont
  {Goltsev}}, \ and\ \bibinfo {author} {\bibfnamefont {J.~F.~F.}\ \bibnamefont
  {Mendes}},\ }\bibfield  {title} {\enquote {\bibinfo {title} {Critical
  phenomena in complex networks},}\ }\href@noop {} {\bibfield  {journal}
  {\bibinfo  {journal} {Rev. Mod. Phys.}\ }\textbf {\bibinfo {volume} {80}},\
  \bibinfo {pages} {1275--1335} (\bibinfo {year} {2008})}\BibitemShut {NoStop}%
\bibitem [{\citenamefont {Arenas}\ \emph {et~al.}(2008)\citenamefont {Arenas},
  \citenamefont {D\'iaz-Guilera}, \citenamefont {Kurths}, \citenamefont
  {Moreno},\ and\ \citenamefont {Zhou}}]{Arenas:2008yq}%
  \BibitemOpen
  \bibfield  {author} {\bibinfo {author} {\bibfnamefont {A.}~\bibnamefont
  {Arenas}}, \bibinfo {author} {\bibfnamefont {A.}~\bibnamefont
  {D\'iaz-Guilera}}, \bibinfo {author} {\bibfnamefont {J.}~\bibnamefont
  {Kurths}}, \bibinfo {author} {\bibfnamefont {Y.}~\bibnamefont {Moreno}}, \
  and\ \bibinfo {author} {\bibfnamefont {C.}~\bibnamefont {Zhou}},\ }\bibfield
  {title} {\enquote {\bibinfo {title} {Synchronization in complex networks},}\
  }\href@noop {} {\bibfield  {journal} {\bibinfo  {journal} {Phys. Rep.}\
  }\textbf {\bibinfo {volume} {469}},\ \bibinfo {pages} {93--153} (\bibinfo
  {year} {2008})}\BibitemShut {NoStop}%
\bibitem [{\citenamefont {D{\"o}rfler}\ and\ \citenamefont
  {Bullo}(2014)}]{Dorfler:2014fk}%
  \BibitemOpen
  \bibfield  {author} {\bibinfo {author} {\bibfnamefont {F.}~\bibnamefont
  {D{\"o}rfler}}\ and\ \bibinfo {author} {\bibfnamefont {F.}~\bibnamefont
  {Bullo}},\ }\bibfield  {title} {\enquote {\bibinfo {title} {Synchronization
  in complex networks of phase oscillators: A survey},}\ }\href@noop {}
  {\bibfield  {journal} {\bibinfo  {journal} {Automatica}\ }\textbf {\bibinfo
  {volume} {50}},\ \bibinfo {pages} {1539--1564} (\bibinfo {year}
  {2014})}\BibitemShut {NoStop}%
\bibitem [{\citenamefont {Mallada}\ and\ \citenamefont
  {Tang}(2011)}]{Mallada:2011lr}%
  \BibitemOpen
  \bibfield  {author} {\bibinfo {author} {\bibfnamefont {E.}~\bibnamefont
  {Mallada}}\ and\ \bibinfo {author} {\bibfnamefont {A.}~\bibnamefont {Tang}},\
  }\bibfield  {title} {\enquote {\bibinfo {title} {Improving damping of power
  networks: Power scheduling and impedance adaptation},}\ }in\ \href@noop {}
  {\emph {\bibinfo {booktitle} {IEEE Conference on Decision and Control}}}\
  (\bibinfo {year} {2011})\ pp.\ \bibinfo {pages} {7729--7734}\BibitemShut
  {NoStop}%
\bibitem [{\citenamefont {Rohden}\ \emph {et~al.}(2012)\citenamefont {Rohden},
  \citenamefont {Sorge}, \citenamefont {Timme},\ and\ \citenamefont
  {Witthaut}}]{PhysRevLett.109.064101}%
  \BibitemOpen
  \bibfield  {author} {\bibinfo {author} {\bibfnamefont {M.}~\bibnamefont
  {Rohden}}, \bibinfo {author} {\bibfnamefont {A.}~\bibnamefont {Sorge}},
  \bibinfo {author} {\bibfnamefont {M.}~\bibnamefont {Timme}}, \ and\ \bibinfo
  {author} {\bibfnamefont {D.}~\bibnamefont {Witthaut}},\ }\bibfield  {title}
  {\enquote {\bibinfo {title} {Self-organized synchronization in decentralized
  power grids},}\ }\href@noop {} {\bibfield  {journal} {\bibinfo  {journal}
  {Phys. Rev. Lett.}\ }\textbf {\bibinfo {volume} {109}},\ \bibinfo {pages}
  {064101} (\bibinfo {year} {2012})}\BibitemShut {NoStop}%
\bibitem [{\citenamefont {Lozano}, \citenamefont {Buzna},\ and\ \citenamefont
  {D{\'\i}az-Guilera}(2012)}]{Lozano:2012qy}%
  \BibitemOpen
  \bibfield  {author} {\bibinfo {author} {\bibfnamefont {S.}~\bibnamefont
  {Lozano}}, \bibinfo {author} {\bibfnamefont {L.}~\bibnamefont {Buzna}}, \
  and\ \bibinfo {author} {\bibfnamefont {A.}~\bibnamefont
  {D{\'\i}az-Guilera}},\ }\bibfield  {title} {\enquote {\bibinfo {title} {Role
  of network topology in the synchronization of power systems},}\ }\href@noop
  {} {\bibfield  {journal} {\bibinfo  {journal} {Eur. Phys. J. B}\ }\textbf
  {\bibinfo {volume} {85}},\ \bibinfo {pages} {1--8} (\bibinfo {year}
  {2012})}\BibitemShut {NoStop}%
\bibitem [{\citenamefont {D{\"o}rfler}, \citenamefont {Chertkov},\ and\
  \citenamefont {Bullo}(2013)}]{Dorfler:2013ve}%
  \BibitemOpen
  \bibfield  {author} {\bibinfo {author} {\bibfnamefont {F.}~\bibnamefont
  {D{\"o}rfler}}, \bibinfo {author} {\bibfnamefont {M.}~\bibnamefont
  {Chertkov}}, \ and\ \bibinfo {author} {\bibfnamefont {F.}~\bibnamefont
  {Bullo}},\ }\bibfield  {title} {\enquote {\bibinfo {title} {Synchronization
  in complex oscillator networks and smart grids},}\ }\href@noop {} {\bibfield
  {journal} {\bibinfo  {journal} {Proc. Natl. Acad. Sci. U.S.A.}\ }\textbf
  {\bibinfo {volume} {110}},\ \bibinfo {pages} {2005--2010} (\bibinfo {year}
  {2013})}\BibitemShut {NoStop}%
\bibitem [{\citenamefont {Motter}\ \emph {et~al.}(2013)\citenamefont {Motter},
  \citenamefont {Myers}, \citenamefont {Anghel},\ and\ \citenamefont
  {Nishikawa}}]{Motter:2013fk}%
  \BibitemOpen
  \bibfield  {author} {\bibinfo {author} {\bibfnamefont {A.~E.}\ \bibnamefont
  {Motter}}, \bibinfo {author} {\bibfnamefont {S.~A.}\ \bibnamefont {Myers}},
  \bibinfo {author} {\bibfnamefont {M.}~\bibnamefont {Anghel}}, \ and\ \bibinfo
  {author} {\bibfnamefont {T.}~\bibnamefont {Nishikawa}},\ }\bibfield  {title}
  {\enquote {\bibinfo {title} {Spontaneous synchrony in power-grid networks},}\
  }\href@noop {} {\bibfield  {journal} {\bibinfo  {journal} {Nat. Phys.}\
  }\textbf {\bibinfo {volume} {9}},\ \bibinfo {pages} {191--197} (\bibinfo
  {year} {2013})}\BibitemShut {NoStop}%
\bibitem [{\citenamefont {Menck}\ \emph {et~al.}(2014)\citenamefont {Menck},
  \citenamefont {Heitzig}, \citenamefont {Kurths},\ and\ \citenamefont
  {Schellnhuber}}]{Menck:2014fk}%
  \BibitemOpen
  \bibfield  {author} {\bibinfo {author} {\bibfnamefont {P.~J.}\ \bibnamefont
  {Menck}}, \bibinfo {author} {\bibfnamefont {J.}~\bibnamefont {Heitzig}},
  \bibinfo {author} {\bibfnamefont {J.}~\bibnamefont {Kurths}}, \ and\ \bibinfo
  {author} {\bibfnamefont {H.~J.}\ \bibnamefont {Schellnhuber}},\ }\bibfield
  {title} {\enquote {\bibinfo {title} {How dead ends undermine power grid
  stability},}\ }\href@noop {} {\bibfield  {journal} {\bibinfo  {journal} {Nat.
  Commun.}\ }\textbf {\bibinfo {volume} {5}},\ \bibinfo {pages} {3969}
  (\bibinfo {year} {2014})}\BibitemShut {NoStop}%
\bibitem [{\citenamefont {Pecora}\ \emph {et~al.}(2014)\citenamefont {Pecora},
  \citenamefont {Sorrentino}, \citenamefont {Hagerstrom}, \citenamefont
  {Murphy},\ and\ \citenamefont {Roy}}]{Pecora:2014zr}%
  \BibitemOpen
  \bibfield  {author} {\bibinfo {author} {\bibfnamefont {L.~M.}\ \bibnamefont
  {Pecora}}, \bibinfo {author} {\bibfnamefont {F.}~\bibnamefont {Sorrentino}},
  \bibinfo {author} {\bibfnamefont {A.~M.}\ \bibnamefont {Hagerstrom}},
  \bibinfo {author} {\bibfnamefont {T.~E.}\ \bibnamefont {Murphy}}, \ and\
  \bibinfo {author} {\bibfnamefont {R.}~\bibnamefont {Roy}},\ }\bibfield
  {title} {\enquote {\bibinfo {title} {Cluster synchronization and isolated
  desynchronization in complex networks with symmetries},}\ }\href@noop {}
  {\bibfield  {journal} {\bibinfo  {journal} {Nat. Commun.}\ }\textbf {\bibinfo
  {volume} {5}},\ \bibinfo {pages} {4079} (\bibinfo {year} {2014})}\BibitemShut
  {NoStop}%
\bibitem [{\citenamefont {Anderson}\ and\ \citenamefont
  {Fouad}(2003)}]{Anderson:1986fk}%
  \BibitemOpen
  \bibfield  {author} {\bibinfo {author} {\bibfnamefont {P.~M.}\ \bibnamefont
  {Anderson}}\ and\ \bibinfo {author} {\bibfnamefont {A.~A.}\ \bibnamefont
  {Fouad}},\ }\href@noop {} {\emph {\bibinfo {title} {Power system control and
  stability}}},\ \bibinfo {edition} {2nd}\ ed.\ (\bibinfo  {publisher} {IEEE
  Press},\ \bibinfo {year} {2003})\BibitemShut {NoStop}%
\bibitem [{\citenamefont {Bergen}\ and\ \citenamefont
  {Hill}(1981)}]{Bergen:1981kx}%
  \BibitemOpen
  \bibfield  {author} {\bibinfo {author} {\bibfnamefont {A.~R.}\ \bibnamefont
  {Bergen}}\ and\ \bibinfo {author} {\bibfnamefont {D.~J.}\ \bibnamefont
  {Hill}},\ }\bibfield  {title} {\enquote {\bibinfo {title} {A structure
  preserving model for power system stability analysis},}\ }\bibfield
  {booktitle} {\emph {\bibinfo {booktitle} {Power Apparatus and Systems, IEEE
  Transactions on}},\ }\href@noop {} {\bibfield  {journal} {\bibinfo  {journal}
  {IEEE Trans. Power Appar. Syst.}\ }\textbf {\bibinfo {volume} {PAS-100}},\
  \bibinfo {pages} {25--35} (\bibinfo {year} {1981})}\BibitemShut {NoStop}%
\bibitem [{\citenamefont {Filatrella}, \citenamefont {Nielsen},\ and\
  \citenamefont {Pedersen}(2008)}]{springerlink:10.1140/epjb/e2008-00098-8}%
  \BibitemOpen
  \bibfield  {author} {\bibinfo {author} {\bibfnamefont {G.}~\bibnamefont
  {Filatrella}}, \bibinfo {author} {\bibfnamefont {A.~H.}\ \bibnamefont
  {Nielsen}}, \ and\ \bibinfo {author} {\bibfnamefont {N.~F.}\ \bibnamefont
  {Pedersen}},\ }\bibfield  {title} {\enquote {\bibinfo {title} {Analysis of a
  power grid using a {K}uramoto-like model},}\ }\href@noop {} {\bibfield
  {journal} {\bibinfo  {journal} {Eur. Phys. J. B}\ }\textbf {\bibinfo {volume}
  {61}},\ \bibinfo {pages} {485--491} (\bibinfo {year} {2008})}\BibitemShut
  {NoStop}%
\bibitem [{\citenamefont {Kuramoto}(1984)}]{kuramoto1984chemical}%
  \BibitemOpen
  \bibfield  {author} {\bibinfo {author} {\bibfnamefont {Y.}~\bibnamefont
  {Kuramoto}},\ }\href@noop {} {\emph {\bibinfo {title} {{Chemical
  oscillations, waves, and turbulence}}}}\ (\bibinfo  {publisher}
  {Springer-Verlag},\ \bibinfo {address} {Berlin},\ \bibinfo {year}
  {1984})\BibitemShut {NoStop}%
\bibitem [{\citenamefont {Tanaka}, \citenamefont {Lichtenberg},\ and\
  \citenamefont {Oishi}(1997)}]{Tanaka1997xd}%
  \BibitemOpen
  \bibfield  {author} {\bibinfo {author} {\bibfnamefont {H.-A.}\ \bibnamefont
  {Tanaka}}, \bibinfo {author} {\bibfnamefont {A.~J.}\ \bibnamefont
  {Lichtenberg}}, \ and\ \bibinfo {author} {\bibfnamefont {S.}~\bibnamefont
  {Oishi}},\ }\bibfield  {title} {\enquote {\bibinfo {title} {First order phase
  transition resulting from finite inertia in coupled oscillator systems},}\
  }\href@noop {} {\bibfield  {journal} {\bibinfo  {journal} {Phys. Rev. Lett.}\
  }\textbf {\bibinfo {volume} {78}},\ \bibinfo {pages} {2104--2107} (\bibinfo
  {year} {1997})}\BibitemShut {NoStop}%
\bibitem [{\citenamefont {Acebr\'on}\ and\ \citenamefont
  {Spigler}(1998)}]{Acebron1998iogw}%
  \BibitemOpen
  \bibfield  {author} {\bibinfo {author} {\bibfnamefont {J.~A.}\ \bibnamefont
  {Acebr\'on}}\ and\ \bibinfo {author} {\bibfnamefont {R.}~\bibnamefont
  {Spigler}},\ }\bibfield  {title} {\enquote {\bibinfo {title} {Adaptive
  frequency model for phase-frequency synchronization in large populations of
  globally coupled nonlinear oscillators},}\ }\href@noop {} {\bibfield
  {journal} {\bibinfo  {journal} {Phys. Rev. Lett.}\ }\textbf {\bibinfo
  {volume} {81}},\ \bibinfo {pages} {2229--2232} (\bibinfo {year}
  {1998})}\BibitemShut {NoStop}%
\bibitem [{\citenamefont {Trees}, \citenamefont {Saranathan},\ and\
  \citenamefont {Stroud}(2005)}]{Trees2005hok}%
  \BibitemOpen
  \bibfield  {author} {\bibinfo {author} {\bibfnamefont {B.~R.}\ \bibnamefont
  {Trees}}, \bibinfo {author} {\bibfnamefont {V.}~\bibnamefont {Saranathan}}, \
  and\ \bibinfo {author} {\bibfnamefont {D.}~\bibnamefont {Stroud}},\
  }\bibfield  {title} {\enquote {\bibinfo {title} {Synchronization in
  disordered {Josephson} junction arrays: Small-world connections and the
  {Kuramoto} model},}\ }\href@noop {} {\bibfield  {journal} {\bibinfo
  {journal} {Phys. Rev. E}\ }\textbf {\bibinfo {volume} {71}},\ \bibinfo
  {pages} {016215} (\bibinfo {year} {2005})}\BibitemShut {NoStop}%
\bibitem [{\citenamefont {Ji}\ \emph {et~al.}(2013)\citenamefont {Ji},
  \citenamefont {Peron}, \citenamefont {Menck}, \citenamefont {Rodrigues},\
  and\ \citenamefont {Kurths}}]{Ji2013xeo}%
  \BibitemOpen
  \bibfield  {author} {\bibinfo {author} {\bibfnamefont {P.}~\bibnamefont
  {Ji}}, \bibinfo {author} {\bibfnamefont {T.~K.~D.}\ \bibnamefont {Peron}},
  \bibinfo {author} {\bibfnamefont {P.~J.}\ \bibnamefont {Menck}}, \bibinfo
  {author} {\bibfnamefont {F.~A.}\ \bibnamefont {Rodrigues}}, \ and\ \bibinfo
  {author} {\bibfnamefont {J.}~\bibnamefont {Kurths}},\ }\bibfield  {title}
  {\enquote {\bibinfo {title} {Cluster explosive synchronization in complex
  networks},}\ }\href@noop {} {\bibfield  {journal} {\bibinfo  {journal} {Phys.
  Rev. Lett.}\ }\textbf {\bibinfo {volume} {110}},\ \bibinfo {pages} {218701}
  (\bibinfo {year} {2013})}\BibitemShut {NoStop}%
\bibitem [{\citenamefont {Grainger}\ and\ \citenamefont {{Stevenson
  Jr.}}(1994)}]{Grainger:1994yo}%
  \BibitemOpen
  \bibfield  {author} {\bibinfo {author} {\bibfnamefont {J.}~\bibnamefont
  {Grainger}}\ and\ \bibinfo {author} {\bibfnamefont {W.}~\bibnamefont
  {{Stevenson Jr.}}},\ }\href@noop {} {\emph {\bibinfo {title} {Power System
  Analysis}}}\ (\bibinfo  {publisher} {McGraw-Hill Co.},\ \bibinfo {address}
  {Singapore},\ \bibinfo {year} {1994})\BibitemShut {NoStop}%
\bibitem [{\citenamefont {Chow}\ and\ \citenamefont
  {Cheung}(1992)}]{chow1992toolbox}%
  \BibitemOpen
  \bibfield  {author} {\bibinfo {author} {\bibfnamefont {J.~H.}\ \bibnamefont
  {Chow}}\ and\ \bibinfo {author} {\bibfnamefont {K.~W.}\ \bibnamefont
  {Cheung}},\ }\bibfield  {title} {\enquote {\bibinfo {title} {A toolbox for
  power system dynamics and control engineering education and research},}\
  }\href@noop {} {\bibfield  {journal} {\bibinfo  {journal} {IEEE T. Power
  Syst.}\ }\textbf {\bibinfo {volume} {7}},\ \bibinfo {pages} {1559--1564}
  (\bibinfo {year} {1992})}\BibitemShut {NoStop}%
\bibitem [{\citenamefont {Zimmerman}, \citenamefont {Murillo-S\'anchez},\ and\
  \citenamefont {Thomas}(2011)}]{matpower}%
  \BibitemOpen
  \bibfield  {author} {\bibinfo {author} {\bibfnamefont {R.~D.}\ \bibnamefont
  {Zimmerman}}, \bibinfo {author} {\bibfnamefont {C.~E.}\ \bibnamefont
  {Murillo-S\'anchez}}, \ and\ \bibinfo {author} {\bibfnamefont {R.~J.}\
  \bibnamefont {Thomas}},\ }\bibfield  {title} {\enquote {\bibinfo {title}
  {{MATPOWER}: Steady-state operations, planning and analysis tools for power
  systems research and education},}\ }\href@noop {} {\bibfield  {journal}
  {\bibinfo  {journal} {IEEE T. Power Syst.}\ }\textbf {\bibinfo {volume}
  {26}},\ \bibinfo {pages} {12--19} (\bibinfo {year} {2011})}\BibitemShut
  {NoStop}%
\bibitem [{\citenamefont {Sauer}\ and\ \citenamefont
  {Pai}(1998)}]{sauer1998power}%
  \BibitemOpen
  \bibfield  {author} {\bibinfo {author} {\bibfnamefont {P.}~\bibnamefont
  {Sauer}}\ and\ \bibinfo {author} {\bibfnamefont {A.}~\bibnamefont {Pai}},\
  }\href@noop {} {\emph {\bibinfo {title} {Power System Dynamics and
  Stability}}}\ (\bibinfo  {publisher} {Prentice Hall},\ \bibinfo {year}
  {1998})\BibitemShut {NoStop}%
\bibitem [{\citenamefont {Caliskan}\ and\ \citenamefont
  {Tabuada}(2014)}]{Caliskan2014slgd}%
  \BibitemOpen
  \bibfield  {author} {\bibinfo {author} {\bibfnamefont {S.}~\bibnamefont
  {Caliskan}}\ and\ \bibinfo {author} {\bibfnamefont {P.}~\bibnamefont
  {Tabuada}},\ }\bibfield  {title} {\enquote {\bibinfo {title} {Compositional
  transient stability analysis of multimachine power networks},}\ }\href@noop
  {} {\bibfield  {journal} {\bibinfo  {journal} {IEEE T. Contr. Netw. Syst.}\
  }\textbf {\bibinfo {volume} {1}},\ \bibinfo {pages} {4--14} (\bibinfo {year}
  {2014})}\BibitemShut {NoStop}%
\bibitem [{\citenamefont {Pai}(1989)}]{pai1989energy}%
  \BibitemOpen
  \bibfield  {author} {\bibinfo {author} {\bibfnamefont {M.}~\bibnamefont
  {Pai}},\ }\href@noop {} {\emph {\bibinfo {title} {Energy function analysis
  for power system stability}}}\ (\bibinfo  {publisher} {Kluwer Academic
  Publishers},\ \bibinfo {address} {Norwell},\ \bibinfo {year}
  {1989})\BibitemShut {NoStop}%
\bibitem [{\citenamefont {Chu}\ and\ \citenamefont
  {Chiang}(1999)}]{Chu:1999kx}%
  \BibitemOpen
  \bibfield  {author} {\bibinfo {author} {\bibfnamefont {C.-C.}\ \bibnamefont
  {Chu}}\ and\ \bibinfo {author} {\bibfnamefont {H.-D.}\ \bibnamefont
  {Chiang}},\ }\bibfield  {title} {\enquote {\bibinfo {title} {Constructing
  analytical energy functions for lossless network-reduction power system
  models: Framework and new developments},}\ }\href@noop {} {\bibfield
  {journal} {\bibinfo  {journal} {Circuits Syst. Signal Process.}\ }\textbf
  {\bibinfo {volume} {18}},\ \bibinfo {pages} {1--16} (\bibinfo {year}
  {1999})}\BibitemShut {NoStop}%
\bibitem [{\citenamefont {Chu}\ and\ \citenamefont
  {Chiang}(2005)}]{Chu:2005uq}%
  \BibitemOpen
  \bibfield  {author} {\bibinfo {author} {\bibfnamefont {C.-C.}\ \bibnamefont
  {Chu}}\ and\ \bibinfo {author} {\bibfnamefont {H.-D.}\ \bibnamefont
  {Chiang}},\ }\bibfield  {title} {\enquote {\bibinfo {title} {Constructing
  analytical energy functions for network-preserving power system models},}\
  }\href@noop {} {\bibfield  {journal} {\bibinfo  {journal} {Circuits Syst.
  Signal Process.}\ }\textbf {\bibinfo {volume} {24}},\ \bibinfo {pages}
  {363--383} (\bibinfo {year} {2005})}\BibitemShut {NoStop}%
\bibitem [{\citenamefont {Tzeng}\ and\ \citenamefont
  {Wu}(2006)}]{Tzeng2006lsg}%
  \BibitemOpen
  \bibfield  {author} {\bibinfo {author} {\bibfnamefont {W.~J.}\ \bibnamefont
  {Tzeng}}\ and\ \bibinfo {author} {\bibfnamefont {F.~Y.}\ \bibnamefont {Wu}},\
  }\bibfield  {title} {\enquote {\bibinfo {title} {Theory of impedance
  networks: the two-point impedance and {LC} resonances},}\ }\href@noop {}
  {\bibfield  {journal} {\bibinfo  {journal} {J. Phys. A-Math. Gen.}\ }\textbf
  {\bibinfo {volume} {39}},\ \bibinfo {pages} {8579} (\bibinfo {year}
  {2006})}\BibitemShut {NoStop}%
\bibitem [{\citenamefont {D\"orfler}\ and\ \citenamefont
  {Bullo}(2013)}]{Dorfler:2011fa}%
  \BibitemOpen
  \bibfield  {author} {\bibinfo {author} {\bibfnamefont {F.}~\bibnamefont
  {D\"orfler}}\ and\ \bibinfo {author} {\bibfnamefont {F.}~\bibnamefont
  {Bullo}},\ }\bibfield  {title} {\enquote {\bibinfo {title} {Kron reduction of
  graphs with applications to electrical networks},}\ }\href@noop {} {\bibfield
   {journal} {\bibinfo  {journal} {IEEE Trans. Circuits Syst. I}\ }\textbf
  {\bibinfo {volume} {60}},\ \bibinfo {pages} {150--163} (\bibinfo {year}
  {2013})}\BibitemShut {NoStop}%
\bibitem [{\citenamefont {Pecora}\ and\ \citenamefont
  {Carroll}(1998)}]{Pecora:1998zp}%
  \BibitemOpen
  \bibfield  {author} {\bibinfo {author} {\bibfnamefont {L.~M.}\ \bibnamefont
  {Pecora}}\ and\ \bibinfo {author} {\bibfnamefont {T.~L.}\ \bibnamefont
  {Carroll}},\ }\bibfield  {title} {\enquote {\bibinfo {title} {Master
  stability functions for synchronized coupled systems},}\ }\href@noop {}
  {\bibfield  {journal} {\bibinfo  {journal} {Phys. Rev. Lett.}\ }\textbf
  {\bibinfo {volume} {80}},\ \bibinfo {pages} {2109--2112} (\bibinfo {year}
  {1998})}\BibitemShut {NoStop}%
\bibitem [{\citenamefont {D\"orfler}\ and\ \citenamefont
  {Bullo}(2012)}]{fd-fb:09z}%
  \BibitemOpen
  \bibfield  {author} {\bibinfo {author} {\bibfnamefont {F.}~\bibnamefont
  {D\"orfler}}\ and\ \bibinfo {author} {\bibfnamefont {F.}~\bibnamefont
  {Bullo}},\ }\bibfield  {title} {\enquote {\bibinfo {title} {Synchronization
  and transient stability in power networks and non-uniform {K}uramoto
  oscillators},}\ }\href@noop {} {\bibfield  {journal} {\bibinfo  {journal}
  {SIAM J. Control Optim.}\ }\textbf {\bibinfo {volume} {50}},\ \bibinfo
  {pages} {1616--1642} (\bibinfo {year} {2012})}\BibitemShut {NoStop}%
\bibitem [{\citenamefont {Wu}(2014)}]{Wu:2014lr}%
  \BibitemOpen
  \bibfield  {author} {\bibinfo {author} {\bibfnamefont {X.}~\bibnamefont
  {Wu}},\ }\bibfield  {title} {\enquote {\bibinfo {title} {Pinning
  synchronization of linear complex coupling synchronous generators network of
  hydroelectric generating set},}\ }\href@noop {} {\bibfield  {journal}
  {\bibinfo  {journal} {Math. Probl. Eng.}\ }\textbf {\bibinfo {volume}
  {2014}},\ \bibinfo {pages} {476794} (\bibinfo {year} {2014})}\BibitemShut
  {NoStop}%
\bibitem [{\citenamefont {Sj\"odin}\ and\ \citenamefont
  {Gayme}(2014)}]{E:2014fk}%
  \BibitemOpen
  \bibfield  {author} {\bibinfo {author} {\bibfnamefont {E.}~\bibnamefont
  {Sj\"odin}}\ and\ \bibinfo {author} {\bibfnamefont {D.~F.}\ \bibnamefont
  {Gayme}},\ }\bibfield  {title} {\enquote {\bibinfo {title} {Transient losses
  in synchronizing renewable energy integrated power networks},}\ }in\
  \href@noop {} {\emph {\bibinfo {booktitle} {American Control Conference}}}\
  (\bibinfo {year} {2014})\ pp.\ \bibinfo {pages} {5217--5223}\BibitemShut
  {NoStop}%
\bibitem [{\citenamefont {Odun-Ayo}\ and\ \citenamefont
  {Crow}(2012)}]{Odun-Ayo:2012qy}%
  \BibitemOpen
  \bibfield  {author} {\bibinfo {author} {\bibfnamefont {T.}~\bibnamefont
  {Odun-Ayo}}\ and\ \bibinfo {author} {\bibfnamefont {M.~L.}\ \bibnamefont
  {Crow}},\ }\bibfield  {title} {\enquote {\bibinfo {title}
  {Structure-preserved power system transient stability using stochastic energy
  functions},}\ }\href@noop {} {\bibfield  {journal} {\bibinfo  {journal} {IEEE
  Trans. Power Syst.}\ }\textbf {\bibinfo {volume} {27}},\ \bibinfo {pages}
  {1450--1458} (\bibinfo {year} {2012})}\BibitemShut {NoStop}%
\bibitem [{\citenamefont {Price}\ \emph {et~al.}(1995)\citenamefont {Price},
  \citenamefont {Taylor}, \citenamefont {Rogers}, \citenamefont {Srinivasan},
  \citenamefont {Concordia}, \citenamefont {Pal}, \citenamefont {Bess},
  \citenamefont {Kundur}, \citenamefont {Agrawal}, \citenamefont {Luini},
  \citenamefont {Vaahedi},\ and\ \citenamefont {Johnson}}]{price1995standard}%
  \BibitemOpen
  \bibfield  {author} {\bibinfo {author} {\bibfnamefont {W.}~\bibnamefont
  {Price}}, \bibinfo {author} {\bibfnamefont {C.}~\bibnamefont {Taylor}},
  \bibinfo {author} {\bibfnamefont {G.}~\bibnamefont {Rogers}}, \bibinfo
  {author} {\bibfnamefont {K.}~\bibnamefont {Srinivasan}}, \bibinfo {author}
  {\bibfnamefont {C.}~\bibnamefont {Concordia}}, \bibinfo {author}
  {\bibfnamefont {M.~K.}\ \bibnamefont {Pal}}, \bibinfo {author} {\bibfnamefont
  {K.~C.}\ \bibnamefont {Bess}}, \bibinfo {author} {\bibfnamefont
  {P.}~\bibnamefont {Kundur}}, \bibinfo {author} {\bibfnamefont {B.~L.}\
  \bibnamefont {Agrawal}}, \bibinfo {author} {\bibfnamefont {J.~F.}\
  \bibnamefont {Luini}}, \bibinfo {author} {\bibfnamefont {E.}~\bibnamefont
  {Vaahedi}}, \ and\ \bibinfo {author} {\bibfnamefont {B.~K.}\ \bibnamefont
  {Johnson}},\ }\bibfield  {title} {\enquote {\bibinfo {title} {Standard load
  models for power flow and dynamic performance simulation},}\ }\href@noop {}
  {\bibfield  {journal} {\bibinfo  {journal} {IEEE T. Power Syst.}\ }\textbf
  {\bibinfo {volume} {10}},\ \bibinfo {pages} {1302--1313} (\bibinfo {year}
  {1995})}\BibitemShut {NoStop}%
\bibitem [{\citenamefont {Witthaut}\ and\ \citenamefont
  {Timme}(2012)}]{Witthaut:2012wd}%
  \BibitemOpen
  \bibfield  {author} {\bibinfo {author} {\bibfnamefont {D.}~\bibnamefont
  {Witthaut}}\ and\ \bibinfo {author} {\bibfnamefont {M.}~\bibnamefont
  {Timme}},\ }\bibfield  {title} {\enquote {\bibinfo {title} {Braess's paradox
  in oscillator networks, desynchronization and power outage},}\ }\href@noop {}
  {\bibfield  {journal} {\bibinfo  {journal} {New J. Phys.}\ }\textbf {\bibinfo
  {volume} {14}},\ \bibinfo {pages} {083036} (\bibinfo {year}
  {2012})}\BibitemShut {NoStop}%
\bibitem [{\citenamefont {Rohden}\ \emph {et~al.}(2014)\citenamefont {Rohden},
  \citenamefont {Sorge}, \citenamefont {Witthaut},\ and\ \citenamefont
  {Timme}}]{Rohden2014ijve}%
  \BibitemOpen
  \bibfield  {author} {\bibinfo {author} {\bibfnamefont {M.}~\bibnamefont
  {Rohden}}, \bibinfo {author} {\bibfnamefont {A.}~\bibnamefont {Sorge}},
  \bibinfo {author} {\bibfnamefont {D.}~\bibnamefont {Witthaut}}, \ and\
  \bibinfo {author} {\bibfnamefont {M.}~\bibnamefont {Timme}},\ }\bibfield
  {title} {\enquote {\bibinfo {title} {Impact of network topology on synchrony
  of oscillatory power grids},}\ }\href@noop {} {\bibfield  {journal} {\bibinfo
   {journal} {Chaos}\ }\textbf {\bibinfo {volume} {24}},\ \bibinfo {eid}
  {013123} (\bibinfo {year} {2014})}\BibitemShut {NoStop}%
\bibitem [{\citenamefont {Buzna}, \citenamefont {Lozano},\ and\ \citenamefont
  {D\'{\i}az-Guilera}(2009)}]{PhysRevE.80.066120}%
  \BibitemOpen
  \bibfield  {author} {\bibinfo {author} {\bibfnamefont {L.}~\bibnamefont
  {Buzna}}, \bibinfo {author} {\bibfnamefont {S.}~\bibnamefont {Lozano}}, \
  and\ \bibinfo {author} {\bibfnamefont {A.}~\bibnamefont
  {D\'{\i}az-Guilera}},\ }\bibfield  {title} {\enquote {\bibinfo {title}
  {Synchronization in symmetric bipolar population networks},}\ }\href@noop {}
  {\bibfield  {journal} {\bibinfo  {journal} {Phys. Rev. E}\ }\textbf {\bibinfo
  {volume} {80}},\ \bibinfo {pages} {066120} (\bibinfo {year}
  {2009})}\BibitemShut {NoStop}%
\bibitem [{\citenamefont {Vittal}(1992)}]{Vittal:1992lr}%
  \BibitemOpen
  \bibfield  {author} {\bibinfo {author} {\bibfnamefont {V.}~\bibnamefont
  {Vittal}},\ }\bibfield  {title} {\enquote {\bibinfo {title} {Transient
  stability test systems for direct stability methods},}\ }\href@noop {}
  {\bibfield  {journal} {\bibinfo  {journal} {IEEE T. Power Syst.}\ }\textbf
  {\bibinfo {volume} {7}},\ \bibinfo {pages} {37--43} (\bibinfo {year}
  {1992})}\BibitemShut {NoStop}%
\bibitem [{\citenamefont {DeMarco}(2001)}]{DeMarco:2001lr}%
  \BibitemOpen
  \bibfield  {author} {\bibinfo {author} {\bibfnamefont {C.~L.}\ \bibnamefont
  {DeMarco}},\ }\bibfield  {title} {\enquote {\bibinfo {title} {A phase
  transition model for cascading network failure},}\ }\href@noop {} {\bibfield
  {journal} {\bibinfo  {journal} {IEEE Contr. Syst. Mag.}\ }\textbf {\bibinfo
  {volume} {21}},\ \bibinfo {pages} {40--51} (\bibinfo {year}
  {2001})}\BibitemShut {NoStop}%
\bibitem [{\citenamefont {Gajduk}, \citenamefont {Todorovski},\ and\
  \citenamefont {Kocarev}(2014)}]{Gajduk:2014kx}%
  \BibitemOpen
  \bibfield  {author} {\bibinfo {author} {\bibfnamefont {A.}~\bibnamefont
  {Gajduk}}, \bibinfo {author} {\bibfnamefont {M.}~\bibnamefont {Todorovski}},
  \ and\ \bibinfo {author} {\bibfnamefont {L.}~\bibnamefont {Kocarev}},\
  }\bibfield  {title} {\enquote {\bibinfo {title} {Stability of power grids: An
  overview},}\ }\href@noop {} {\bibfield  {journal} {\bibinfo  {journal} {Eur.
  Phys. J.}\ }\textbf {\bibinfo {volume} {223}},\ \bibinfo {pages} {2387--2409}
  (\bibinfo {year} {2014})}\BibitemShut {NoStop}%
\bibitem [{\citenamefont {Menck}\ \emph {et~al.}(2013)\citenamefont {Menck},
  \citenamefont {Heitzig}, \citenamefont {Marwan},\ and\ \citenamefont
  {Kurths}}]{Menck:2013fk}%
  \BibitemOpen
  \bibfield  {author} {\bibinfo {author} {\bibfnamefont {P.~J.}\ \bibnamefont
  {Menck}}, \bibinfo {author} {\bibfnamefont {J.}~\bibnamefont {Heitzig}},
  \bibinfo {author} {\bibfnamefont {N.}~\bibnamefont {Marwan}}, \ and\ \bibinfo
  {author} {\bibfnamefont {J.}~\bibnamefont {Kurths}},\ }\bibfield  {title}
  {\enquote {\bibinfo {title} {How basin stability complements the
  linear-stability paradigm},}\ }\href@noop {} {\bibfield  {journal} {\bibinfo
  {journal} {Nat. Phys.}\ }\textbf {\bibinfo {volume} {9}},\ \bibinfo {pages}
  {89--92} (\bibinfo {year} {2013})}\BibitemShut {NoStop}%
\bibitem [{\citenamefont {Mezi{\'c}}(2005)}]{Mezic:2005fj}%
  \BibitemOpen
  \bibfield  {author} {\bibinfo {author} {\bibfnamefont {I.}~\bibnamefont
  {Mezi{\'c}}},\ }\bibfield  {title} {\enquote {\bibinfo {title} {Spectral
  properties of dynamical systems, model reduction and decompositions},}\
  }\href@noop {} {\bibfield  {journal} {\bibinfo  {journal} {Nonlinear Dynam.}\
  }\textbf {\bibinfo {volume} {41}},\ \bibinfo {pages} {309--325} (\bibinfo
  {year} {2005})}\BibitemShut {NoStop}%
\bibitem [{\citenamefont {Susuki}\ and\ \citenamefont
  {Mezi{\'c}}(2011)}]{Susuki:2011qy}%
  \BibitemOpen
  \bibfield  {author} {\bibinfo {author} {\bibfnamefont {Y.}~\bibnamefont
  {Susuki}}\ and\ \bibinfo {author} {\bibfnamefont {I.}~\bibnamefont
  {Mezi{\'c}}},\ }\bibfield  {title} {\enquote {\bibinfo {title} {Nonlinear
  {Koopman} modes and coherency identification of coupled swing dynamics},}\
  }\href@noop {} {\bibfield  {journal} {\bibinfo  {journal} {IEEE T. Power
  Syst.}\ }\textbf {\bibinfo {volume} {26}},\ \bibinfo {pages} {1894--1904}
  (\bibinfo {year} {2011})}\BibitemShut {NoStop}%
\bibitem [{\citenamefont {Nishikawa}\ and\ \citenamefont
  {Motter}(2010)}]{Nishikawa:2010fk}%
  \BibitemOpen
  \bibfield  {author} {\bibinfo {author} {\bibfnamefont {T.}~\bibnamefont
  {Nishikawa}}\ and\ \bibinfo {author} {\bibfnamefont {A.~E.}\ \bibnamefont
  {Motter}},\ }\bibfield  {title} {\enquote {\bibinfo {title} {Network
  synchronization landscape reveals compensatory structures, quantization, and
  the positive effect of negative interactions},}\ }\href@noop {} {\bibfield
  {journal} {\bibinfo  {journal} {Proc. Natl. Acad. Sci. U.S.A.}\ }\textbf
  {\bibinfo {volume} {107}},\ \bibinfo {pages} {10342--10347} (\bibinfo {year}
  {2010})}\BibitemShut {NoStop}%
\bibitem [{\citenamefont {Beno\^{i}t}(1991)}]{E:1991lr}%
  \BibitemOpen
  \bibfield  {author} {\bibinfo {author} {\bibfnamefont {E.}~\bibnamefont
  {Beno\^{i}t}},\ }\href@noop {} {\emph {\bibinfo {title} {Dynamic
  Bifurcations}}},\ \bibinfo {series} {Lecture Notes in Mathematics}, Vol.\
  \bibinfo {volume} {1493}\ (\bibinfo  {publisher} {Springer Berlin
  Heidelberg},\ \bibinfo {year} {1991})\BibitemShut {NoStop}%
\bibitem [{\citenamefont {Dobson}(1992)}]{Dobson1992kls}%
  \BibitemOpen
  \bibfield  {author} {\bibinfo {author} {\bibfnamefont {I.}~\bibnamefont
  {Dobson}},\ }\bibfield  {title} {\enquote {\bibinfo {title} {Observations on
  the geometry of saddle node bifurcation and voltage collapse in electrical
  power systems},}\ }\href@noop {} {\bibfield  {journal} {\bibinfo  {journal}
  {IEEE Trans. Circuits Syst. I}\ }\textbf {\bibinfo {volume} {39}},\ \bibinfo
  {pages} {240--243} (\bibinfo {year} {1992})}\BibitemShut {NoStop}%
\bibitem [{\citenamefont {Chiang}\ \emph {et~al.}(1990)\citenamefont {Chiang},
  \citenamefont {Dobson}, \citenamefont {Thomas}, \citenamefont {Thorp},\ and\
  \citenamefont {Fekih-Ahmed}}]{Chiang1990kgs}%
  \BibitemOpen
  \bibfield  {author} {\bibinfo {author} {\bibfnamefont {H.-D.}\ \bibnamefont
  {Chiang}}, \bibinfo {author} {\bibfnamefont {I.}~\bibnamefont {Dobson}},
  \bibinfo {author} {\bibfnamefont {R.}~\bibnamefont {Thomas}}, \bibinfo
  {author} {\bibfnamefont {J.}~\bibnamefont {Thorp}}, \ and\ \bibinfo {author}
  {\bibfnamefont {L.}~\bibnamefont {Fekih-Ahmed}},\ }\bibfield  {title}
  {\enquote {\bibinfo {title} {On voltage collapse in electric power
  systems},}\ }\href@noop {} {\bibfield  {journal} {\bibinfo  {journal} {IEEE
  T. Power Syst.}\ }\textbf {\bibinfo {volume} {5}},\ \bibinfo {pages}
  {601--611} (\bibinfo {year} {1990})}\BibitemShut {NoStop}%
\bibitem [{\citenamefont {Schmietendorf}\ \emph {et~al.}(2014)\citenamefont
  {Schmietendorf}, \citenamefont {Peinke}, \citenamefont {Friedrich},\ and\
  \citenamefont {Kamps}}]{Schmietendorf:2014ys}%
  \BibitemOpen
  \bibfield  {author} {\bibinfo {author} {\bibfnamefont {K.}~\bibnamefont
  {Schmietendorf}}, \bibinfo {author} {\bibfnamefont {J.}~\bibnamefont
  {Peinke}}, \bibinfo {author} {\bibfnamefont {R.}~\bibnamefont {Friedrich}}, \
  and\ \bibinfo {author} {\bibfnamefont {O.}~\bibnamefont {Kamps}},\ }\bibfield
   {title} {\enquote {\bibinfo {title} {Self-organized synchronization and
  voltage stability in networks of synchronous machines},}\ }\href@noop {}
  {\bibfield  {journal} {\bibinfo  {journal} {Eur. Phys. J.}\ }\textbf
  {\bibinfo {volume} {223}},\ \bibinfo {pages} {2577--2592} (\bibinfo {year}
  {2014})}\BibitemShut {NoStop}%
\bibitem [{\citenamefont {Nagata}\ \emph {et~al.}(2014)\citenamefont {Nagata},
  \citenamefont {Fujiwara}, \citenamefont {Tanaka}, \citenamefont {Suzuki},
  \citenamefont {Kohda},\ and\ \citenamefont {Aihara}}]{Nagata:2014vn}%
  \BibitemOpen
  \bibfield  {author} {\bibinfo {author} {\bibfnamefont {M.}~\bibnamefont
  {Nagata}}, \bibinfo {author} {\bibfnamefont {N.}~\bibnamefont {Fujiwara}},
  \bibinfo {author} {\bibfnamefont {G.}~\bibnamefont {Tanaka}}, \bibinfo
  {author} {\bibfnamefont {H.}~\bibnamefont {Suzuki}}, \bibinfo {author}
  {\bibfnamefont {E.}~\bibnamefont {Kohda}}, \ and\ \bibinfo {author}
  {\bibfnamefont {K.}~\bibnamefont {Aihara}},\ }\bibfield  {title} {\enquote
  {\bibinfo {title} {Node-wise robustness against fluctuations of power
  consumption in power grids},}\ }\href@noop {} {\bibfield  {journal} {\bibinfo
   {journal} {Eur. Phys. J.}\ }\textbf {\bibinfo {volume} {223}},\ \bibinfo
  {pages} {2549--2559} (\bibinfo {year} {2014})}\BibitemShut {NoStop}%
\bibitem [{\citenamefont {Cotilla-Sanchez}, \citenamefont {Hines},\ and\
  \citenamefont {Danforth}(2012)}]{Cotilla-Sanchez:2012fk}%
  \BibitemOpen
  \bibfield  {author} {\bibinfo {author} {\bibfnamefont {E.}~\bibnamefont
  {Cotilla-Sanchez}}, \bibinfo {author} {\bibfnamefont {P.~D.~H.}\ \bibnamefont
  {Hines}}, \ and\ \bibinfo {author} {\bibfnamefont {C.~M.}\ \bibnamefont
  {Danforth}},\ }\bibfield  {title} {\enquote {\bibinfo {title} {Predicting
  critical transitions from time series synchrophasor data},}\ }\href@noop {}
  {\bibfield  {journal} {\bibinfo  {journal} {IEEE Trans. Smart Grid}\ }\textbf
  {\bibinfo {volume} {3}},\ \bibinfo {pages} {1832--1840} (\bibinfo {year}
  {2012})}\BibitemShut {NoStop}%
\bibitem [{\citenamefont {Anghel}, \citenamefont {Werley},\ and\ \citenamefont
  {Motter}(2007)}]{Anghel:2007uq}%
  \BibitemOpen
  \bibfield  {author} {\bibinfo {author} {\bibfnamefont {M.}~\bibnamefont
  {Anghel}}, \bibinfo {author} {\bibfnamefont {K.~A.}\ \bibnamefont {Werley}},
  \ and\ \bibinfo {author} {\bibfnamefont {A.~E.}\ \bibnamefont {Motter}},\
  }\bibfield  {title} {\enquote {\bibinfo {title} {Stochastic model for power
  grid dynamics},}\ }in\ \href@noop {} {\emph {\bibinfo {booktitle} {Fortieth
  Hawaii International Conference on System Sciences}}}\ (\bibinfo {address}
  {Big Island, Hawaii},\ \bibinfo {year} {2007})\ p.\ \bibinfo {pages}
  {113}\BibitemShut {NoStop}%
\bibitem [{\citenamefont {Nardelli}\ \emph {et~al.}(2014)\citenamefont
  {Nardelli}, \citenamefont {Rubido}, \citenamefont {Wang}, \citenamefont
  {Baptista}, \citenamefont {Pomalaza-Raez}, \citenamefont {Cardieri},\ and\
  \citenamefont {Latva-aho}}]{Nardelli:2014uq}%
  \BibitemOpen
  \bibfield  {author} {\bibinfo {author} {\bibfnamefont {P.~H.~J.}\
  \bibnamefont {Nardelli}}, \bibinfo {author} {\bibfnamefont {N.}~\bibnamefont
  {Rubido}}, \bibinfo {author} {\bibfnamefont {C.}~\bibnamefont {Wang}},
  \bibinfo {author} {\bibfnamefont {M.~S.}\ \bibnamefont {Baptista}}, \bibinfo
  {author} {\bibfnamefont {C.}~\bibnamefont {Pomalaza-Raez}}, \bibinfo {author}
  {\bibfnamefont {P.}~\bibnamefont {Cardieri}}, \ and\ \bibinfo {author}
  {\bibfnamefont {M.}~\bibnamefont {Latva-aho}},\ }\bibfield  {title} {\enquote
  {\bibinfo {title} {Models for the modern power grid},}\ }\href@noop {}
  {\bibfield  {journal} {\bibinfo  {journal} {Eur. Phys. J.}\ }\textbf
  {\bibinfo {volume} {223}},\ \bibinfo {pages} {2423--2437} (\bibinfo {year}
  {2014})}\BibitemShut {NoStop}%
\end{thebibliography}%

\end{document}